\pdfoutput=1
\documentclass[atoms,article,submit,pdftex,moreauthors]{mdpi} 
\renewcommand{\linenumbers}{}
\firstpage{1} 
\pubvolume{1}
\issuenum{1}
\articlenumber{0}
\pubyear{2025}
\copyrightyear{2025}
\datereceived{ } 
\daterevised{ } 
\dateaccepted{ } 
\datepublished{ } 
\hreflink{https://doi.org/} 

\usepackage{longtable}

\DeclareRobustCommand{\ion}[2]{\textup{#1\,\textsc{\lowercase{#2}}}}

\newcommand{\lsiv}{LS\,IV\-$-$14$^\circ$\-116}
\newcommand{\EC}{EC\,22536\-$-$5304}

\Title{Radiative Data for As III, Se III, Hf IV, and Tl IV: HFR and MCDHF Calculations for Hot Subdwarf Spectroscopy}

\TitleCitation{Title}

\Author{Jérôme Deprince $^{1,2,}$*\orcidA{}, Lucas Maison $^{1}$, Solène Beauraind $^{1}$, Hélène Dupuis $^{1}$, Matti Dorsch $^{3}$ and Pascal Quinet $^{1,4}$}

\AuthorNames{Firstname Lastname, Firstname Lastname and Firstname Lastname}

\isAPAStyle{%
       \AuthorCitation{Lastname, F., Lastname, F., \& Lastname, F.}
         }{%
        \isChicagoStyle{%
        \AuthorCitation{Lastname, Firstname, Firstname Lastname, and Firstname Lastname.}
        }{
        \AuthorCitation{Lastname, F.; Lastname, F.; Lastname, F.}
        }
}

\address{%
$^{1}$ \quad Atomic Physics and Astrophysics, Université de Mons (UMONS), 7000 Mons, Belgium\\
$^{2}$ \quad Astronomy and Astrophysics Institute, Université Libre de Bruxelles (ULB), 1040 Brussels, Belgium\\
$^{3}$ \quad Institut für Physik und Astronomie, Universität Potsdam, 14476 Potsdam, Germany\\
$^{4}$ \quad IPNAS, Université de Liège (ULiège), 4000 Liège, Belgium 
}

\corres{Correspondence: Jerome.Deprince@umons.ac.be}

\abstract{Accurate oscillator strengths for doubly and triply ionized heavy elements are required for quantitative spectroscopy of hot subdwarfs, but available data remain sparse. We present new radiative data for \ion{As}{iii}, \ion{Se}{iii}, \ion{Hf}{iv}, and \ion{Tl}{iv}, calculated with the pseudo-relativistic Hartree--Fock (HFR) method including core-polarization effects and semi-empirical adjustment of radial parameters to experimental energy levels. Independent multiconfiguration Dirac--Hartree--Fock (MCDHF) computations were also performed for \ion{As}{iii}, \ion{Hf}{iv}, and \ion{Tl}{iv}, while published MCDHF data were adopted for \ion{Se}{iii} to assess the reliability of the HFR oscillator strengths. The largest discrepancies between the two approaches are mainly associated with weak transitions, strong cancellation effects, or significant gauge disagreement. For strong transitions satisfying the adopted reliability criteria, we found a good agreement between the oscillator strengths obtained by both methods. The new data have already been applied to the spectra of the heavy-metal hot subdwarfs \lsiv\ and \EC, enabling abundance determinations for As, Se, Hf, and Tl and providing additional lines for future spectroscopic studies.}


\begin{document}

\section{Introduction}


Heavy elements beyond the iron group are predominantly produced by neutron-capture nucleosynthesis, and their observed abundances provide key constraints on the physical sites and conditions of these processes. 
In particular, slow neutron capture (the s-process) is understood to operate in asymptotic giant branch (AGB) stars \citep{Kaeppeler2011}, a conclusion supported by spectroscopic detections of heavy-metal absorption lines in AGB stellar atmospheres, in stars that have accreted material from former AGB companions, and in stars formed from enriched material. Quantitative abundance analyses of such objects play a central role in testing models of stellar evolution and nucleosynthesis \citep{Karakas2014}, but they rely critically on accurate atomic data, including energy levels, transition wavelengths, and oscillator strengths, for the relevant heavy elements. 

Extensive and reliable atomic data are available for neutral and singly ionised heavy metals, sufficient for the analysis of most cool to warm stars, from M-type giants to heavy-metal polluted F-type stars, as well as cool post-AGB stars where photospheric abundances may reflect earlier s-process enrichment \citep[e.g.][]{DeSmedt2012_pAGB_sprocess}. At the other extreme, significant effort has been made to provide data for ionisation stages \textsc{v–vii}, enabling heavy-element studies in very hot white dwarfs \citep{Chayer2005, Rauch2012_Ge, Rauch2019_summary}. However, a substantial gap remains for doubly and triply ionised species, which dominate the spectra of stars with effective temperatures between approximately 20\,000 and 40\,000\,K. 

This temperature regime encompasses the helium-burning subdwarf B (sdB) stars \citep[][for a review]{heber2024}  and extends from early B-type to mid-O-type main-sequence stars. In sdB atmospheres, atomic diffusion can produce overabundances of up to 3\,dex relative to solar values for elements from gallium to lead, predominantly in ionisation stages \textsc{iii–iv} \citep{O'Toole2006b, Chayer2006}. Stage \textsc{iii} lines also appear in cooler chemically peculiar stars: blue horizontal branch stars and HgMn stars, where diffusion-driven abundance anomalies are well established \citep{Michaud1970, Michaud2008}.


The most extreme heavy-element enrichments are found in a group of helium-rich OB-type subdwarf stars (He-sdOB), known as ``heavy-metal'' hot subdwarfs. The prototype of this class is \lsiv, which exhibits 10,000-fold overabundances of zirconium, strontium, and yttrium relative to solar values \citep{Naslim2011}. Other members are instead strongly enriched in lead; a prominent example is \EC, with a lead abundance exceeding one million times solar \citep{Dorsch2021}. While these enrichments were initially attributed to diffusion, recent evolutionary models demonstrate that strong heavy-element enrichment in stars like \EC\ may instead be self-synthesised \citep{Battich2023, Battich2025}, produced by intermediate (i-) neutron-capture processes operating during the helium shell flashes experienced by these core helium-burning stars.

Our work is motivated by the detection of \ion{As}{iii} and \ion{Se}{iii} lines in the optical spectrum of \lsiv\ \citep{Dorsch2020}, as well as \ion{Hf}{iv} and \ion{Tl}{iv} lines in the ultraviolet (UV) spectrum of \EC.
While experimental energy levels and line positions exist for these ions, oscillator strengths have not previously been determined (besides \ion{Se}{iii} \citep{Kitoviene2024_SeIIIBrIVKrV}). We present calculations that fill this gap, enabling abundance measurements for four additional heavy elements in heavy-metal hot subdwarfs. For this purpose, we use the pseudo-relativistic Hartree--Fock (HFR) method including core-polarization effects and semi-empirical adjustment of radial parameters to experimental energy levels. In addition, we also perform independent multiconfiguration Dirac--Hartree--Fock (MCDHF) computations (except for \ion{Se}{iii}, for which recent accurate MCDHF calculations already exist \cite{Kitoviene2024_SeIIIBrIVKrV}) to assess the reliability of the oscillator strengths calculated with HFR. Our results are applied in a parallel paper \citep{Dorsch2026}, which presents the analysis of UV spectra of \lsiv\ and \EC\ obtained with the Hubble Space Telescope (HST).
The same atomic data may prove valuable for analysing other classes of chemically peculiar stars where these ions are present.

\section{Theoretical approaches}

\subsection{The pseudo-relativistic Hartree-Fock method (HFR)}
The atomic structure calculations were performed using the HFR method, as implemented in Cowan’s suite of codes \citep{Cowan1981}. Within this framework, one-electron orbitals are determined separately for each configuration included in the model by solving the coupled Hartree--Fock equations obtained from a variational minimization of the configuration-average energy. The resulting integro-differential equations are treated self-consistently. Since the orbitals associated with different configurations are optimized independently, the resulting non-orthogonality between them is neglected.

Some relativistic effects are included perturbatively: the Blume--Watson spin--orbit interaction (including the one-body Breit operator), the mass--velocity, as well as one-body Darwin corrections.

Following the Slater--Condon formalism, the atomic wavefunctions, i.e. the eigenfunctions of the Hamiltonian, are expressed as linear combinations of configuration state functions (CSFs) in the $LSJ\pi$ coupling scheme,
\begin{equation}
\Psi(\gamma J M_J \pi)= \sum_i^{N_{\texttt{CSF}}} c_i \, \phi(\gamma_i L_i S_i J M_J \pi).
\end{equation}

The multiconfiguration Hamiltonian matrix is constructed and diagonalized using the Slater--Condon theory, with matrix elements written as
\begin{equation}
\langle i | H | j \rangle = \sum_l v^l_{ij} \, x_l,
\end{equation}
where $v^l_{ij}$ denote Racah angular coefficients and $x_l$ the corresponding radial Slater and spin--orbit integrals. Before the semi-empirical least-squares optimization, the electrostatic Slater integrals $F^k$ and $G^k$ (and, when applicable, configuration-interaction integrals $R^k$) were scaled to 85\% of their ab initio HFR values, following the usual Cowan-code prescription \citep{Cowan1981}. This empirical starting-point scaling partially compensates for correlation effects arising from configurations not explicitly included in the model. The 0.85 factor is not applied to the level energies themselves, nor is it used to rescale the spin--orbit parameters.

Core–valence correlation effects are not fully accounted for by the configuration interaction expansion within the HFR method. To compensate for this limitation, the Cowan's code has been modified to take core-polarization (CPOL) effects into account \citep[HFR+CPOL, see][for more details]{Quinet1999,Quinet2002}. In addition, this approach can also be combined with a semi-empirical least-squares fit of radial parameters to minimize the differences between computed and available experimental energy levels.

Radiative transition energies and oscillator strengths for all electric-dipole allowed transitions are subsequently obtained from the eigenvalues and eigenvectors produced by the diagonalization procedure.

The degree of cancellation affecting the calculated transition parameters can be assessed through the cancellation factor (CF), as defined by Cowan \cite{Cowan1981}. For an electric-dipole transition between two states $\Psi_i$ and $\Psi_j$, the line strength is given by
\begin{equation}
    S_{ij} =
    \left|
    \left\langle \Psi_j \left\| \mathbf{P}^{(1)} \right\| \Psi_i \right\rangle
    \right|^2,
\end{equation}
where $\mathbf{P}^{(1)}$ is the electric-dipole transition operator. Expanding the atomic wavefunctions over basis states, the cancellation factor can be written as
\begin{equation}
    \mathrm{CF}_{ij} =
    \left(
    \frac{
    \left|
    \displaystyle\sum_b \sum_c
    y_j^b y_i^c
    \left\langle \psi_c \left\| \mathbf{P}^{(1)} \right\| \psi_b \right\rangle
    \right|
    }{
    \displaystyle\sum_b \sum_c
    \left|
    y_j^b y_i^c
    \left\langle \psi_c \left\| \mathbf{P}^{(1)} \right\| \psi_b \right\rangle
    \right|
    }
    \right)^2 ,
\end{equation}
where $y_i^c$ and $y_j^b$ are the expansion coefficients of the initial and final states, respectively. Small CF values (close to zero) indicate strong cancellation effects between different contributions of opposite sign to the transition matrix element and therefore a potentially less reliable oscillator strength.

\subsection{The multiconfiguration Dirac-Hartree-Fock method (MCDHF)}
The MCDHF method, developed by \citet{Grant2007}, provides a fully relativistic description of atomic systems in which electron correlation is treated through a multiconfiguration expansion of the atomic state functions (ASFs). For a state with total angular momentum $J$, projection $M_J$, and parity $P$, the ASF is expressed as a linear combination of configuration state functions (CSFs),
\begin{equation}
    \Psi(\Gamma PJM_J)=\sum_{i=1}^{N_{\mathrm{CSF}}}c_i\Phi(\gamma_i PJM_J),
\end{equation}
where $\gamma_i$ represents all quantum numbers needed to represent the CSF uniquely and $c_i$ corresponds to the expansion coefficients. The CSFs are antisymmetrized many-electron functions constructed from a common set of orthonormal relativistic one-electron orbitals. In the $jj$-coupling representation used in GRASP \citep{GRASP_manual}, each orbital is described by a four-component Dirac spinor,
\begin{equation}
\psi_{n\kappa m}(\mathbf{r})=\frac{1}{r}
\begin{pmatrix}
P_{n \kappa}(r)\Omega_{\kappa m}(\theta,\phi)\\
iQ_{n \kappa}(r)\Omega_{-\kappa m}(\theta,\phi)\\
\end{pmatrix},
\end{equation}
where $P_{n\kappa}(r)$ and $Q_{n\kappa}(r)$ are the large and small radial components, respectively, and $\Omega_{\kappa m}$ denotes a spherical spinor. A self-consistent field procedure is then used to iteratively optimize the radial components and the expansion coefficients in the extended optimal level scheme (EOL). The optimized orbital basis is subsequently used in a relativistic configuration interaction (RCI) calculation. In this step, the orbitals are kept fixed and the CSF expansion coefficients are obtained by diagonalizing the Hamiltonian matrix. The Hamiltonian can be extended at this stage to include the frequency-independent Breit interaction as well as QED corrections such as vacuum polarization and self-energy effects. Finally, the resulting ASFs are used to evaluate the radiative E1 transition parameters through the reduced matrix elements of the electric dipole transition operator $\mathbf{T^{(1)}}$, $\left\langle \Psi_f \left\| \mathbf{T^{(1)}} \right\| \Psi_i \right\rangle$, between the initial and final states from which the transition rate and other radiative parameters can be determined. The E1 transition matrix elements are calculated in both the Babushkin and Coulomb gauges, corresponding respectively to the length and velocity forms in the non-relativistic limit. Due to the approximate character of the calculated ASFs, differences arise between the results obtained in the two gauges. Therefore an indication of the quality of the calculated transition rates can be provided by the $dT$ parameter \citep{Ekman2014}, which quantifies the relative difference between the two gauges,
\begin{equation}
dT=\frac{|A_{\mathrm{B}}-A_{\mathrm{C}}|}
{\max(A_{\mathrm{B}},A_{\mathrm{C}})},
\end{equation}
with $A_{\mathrm{B}}$ and $A_{\mathrm{C}}$ representing the transition probabilities in the Babushkin and Coulomb gauge, respectively.


\section{Atomic models and computational approaches}

\subsection{HFR models}
The multiconfiguration models used for the four ions considered in this work are described in this section. For all of them, the core-polarization correction has been included and a semi-empirical adjustment procedure has been carried out. The details are given below for each ion, and the comparison between the ab initio and fitted values of the HFR parameters are given in Appendix \ref{app:hfrfit}.\\ 
\textbf{\ion{As}{iii}.} For \ion{As}{iii}, the atomic structure calculations were carried out using an extended multiconfiguration expansion primarily designed to describe valence correlation effects within the HFR framework. The configuration interaction expansion includes 4s$^2$\{4p, 4d, 5s, 5p, 5d, 5f, 6s, 6p, 6d, 6f\}, 4s4p\{4d, 4f, 5s, 5p, 5d, 5f, 6s, 6p, 6d, 6f\}, 4s4d\{4f, 5s\}, 4s\{4p$^2$, 4d$^2$, 4f$^2$\}, 4p\{4d$^2$, 4f$^2$\}, 4p$^2$4d, and 4p$^3$. This model was inspired by the one used by \citet{Rauch2016b} for the isoelectronic ion \ion{Kr}{vi}.
Since core--valence correlation effects are not fully accounted for within the HFR approach, they were treated through the inclusion of a core--polarization correction. An \ion{As}{vi} ionic core was adopted, with a dipole polarizability $\alpha_d = 0.54$ $a_0^3$ taken from \citet{Fraga1976} and a cut-off radius $r_c = 0.66$ $a_0$. The least-squares fit used 21 observed levels belonging to the 4s$^2$4p, 4s$^2$5p, 4s$^2$4f, 4p$^3$, 4s4p$^2$, 4s$^2$5s, 4s$^2$6s, 4s$^2$4d, 4s$^2$5d, and 4s$^2$5g configurations, based on the experimental compilation in the NIST Atomic Spectra Database, on the measurements of \citet{Lang1928} and on the revised analysis carried out by \citet{Churilov96}. The fitted parameters are listed in Table~\ref{tab:as-fit-params}, while the fitted level energies are given in Table~\ref{tab:as-fit-levels}. The mean absolute relative deviation of the fitted levels decreases from 4.2\% in the ab-initio calculation to about 0.01\% after the adjustment. The improvement is especially pronounced for the strongly displaced 4p$^3$ levels and for several even-parity 4s4p$^2$ and 4s$^2n$d levels. The fitted electrostatic integrals associated with 4s4p$^2$ remain within about 5--15\% of their unscaled HFR values, whereas the spin--orbit parameters of the 4d and 5d orbitals require larger corrections. Such effective-parameter changes should not be interpreted independently as direct physical observables; rather, they compensate for correlation contributions missing from the finite configuration-interaction expansion. The resulting improvement in both level positions and configuration mixing provides a more appropriate wavefunction basis for the final HFR+CPOL radiative calculations.

\textbf{\ion{Se}{iii}.} The atomic model adopted for \ion{Se}{iii} is similar to that adopted by \citet{Rauch2016b} for the isoelectronic ion \ion{Kr}{v} and is based on a large multiconfiguration expansion aimed at representing valence correlation effects in this ion. The configuration set comprises
4s$^2$4p$^2$, 4s$^2$4p\{4d, 4f, 5s, 5p, 5d, 5f, 5g, 6s, 6p, 6d, 6f, 7s, 7d, 8s\}, 4s4p$^2$\{4d, 4f, 5s, 5p, 5d, 5f, 6s, 6p, 6d, 6f\}, 4s\{4p4d4f, 4p4d5s, 4p4d$^2$, 4p4f$^2$, 4p$^3$\}, 4p$^2$\{4d$^2$, 4f$^2$\}, 4p$^3$\{4d, 4f, 5s\}, and 4p$^4$.
For the core--polarization potential, an \ion{Se}{vii} ionic core was used, with a dipole polarizability $\alpha_d = 0.36$ $a_0^3$ from \citet{Johnson1983} and a cut-off radius $r_c = 0.62$ $a_0$. The semi-empirical optimization was performed on 85 experimental levels. Most energies were taken from \citet{Tauheed2012}; the set was supplemented with seven levels of the 4s$^2$4p5p configuration from the NIST Atomic Spectra Database \citep{NIST_ASD}. The optimized parameters are listed in Table~\ref{tab:se-fit-params} and the fitted levels in Table~\ref{tab:se-fit-levels}. The mean absolute relative deviation is reduced from 2.91\% before the least-squares optimization to approximately 0.04\% after the fit. Most fitted spin--orbit parameters remain relatively close to their HFR values, although the 4d spin--orbit parameter is increased by approximately a factor of two. Larger changes occur for some exchange integrals of the excited 4s$^2$4p$nl$ configurations, most notably for $G^2$(4p,5p), reflecting the sensitivity of these highly mixed states to correlation effects not explicitly represented by the configuration expansion. 

\textbf{\ion{Hf}{iv}.} For the heavy ion \ion{Hf}{iv}, the atomic structure was modeled using an extensive multiconfiguration expansion following the work of \citet{Quinet2004} for the isoelectronic ion \ion{Lu}{iii}. The configuration interaction expansion includes 4f$^{14}$\{5d, 6s, 6p, 6d, 6f, 6g, 7s, 7p, 7d, 7f, 7g, 8s, 8p, 8d, 8f, 8g, 9s, 9p, 9d, 9f, 9g, 10s, 10p, 10d, 10f, 10g\}, and 4f$^{13}$\{5d$^2$, 5d6s, 5d6p, 6s$^2$, 6s6p\}. As for the CPOL effect modeling, an \ion{Hf}{vi} ionic core was adopted, with a dipole polarizability $\alpha_d = 3.48$ $a_0^3$, extrapolated from the data of \citet{Fraga1976}, and a cut-off radius $r_c = 0.63$ $a_0$. The least-squares optimization used 38 experimental levels reported by \citet{Klinkenberg1961} and \citet{Sugar1974_HfIV}; these levels are listed in Table~\ref{tab:hf-fit-levels}. The mean absolute relative deviation is reduced from 2.15\% in the ab-initio calculation to 0.37\% after optimization. The gain is particularly strong for the relatively pure levels belonging to 4f$^{14}n$s, 4f$^{14}n$d, 4f$^{14}n$p, and 4f$^{14}n$f. The residual discrepancies are concentrated in the strongly mixed 4f$^{13}$5d$^2$ configuration: several of these levels remain at the 0.5--1.8\% level after the fit, and a few individual levels that were already close to experiment before the adjustment become slightly less accurate. This behavior is expected for a global least-squares fit of a dense and strongly interacting structure and emphasizes that the quality of the fit should be assessed globally, rather than level by level.

\textbf{\ion{Tl}{iv}.} The atomic structure of \ion{Tl}{iv} was described using a multiconfiguration expansion based on the model used for the isoelectronic ion \ion{Au}{ii} by \citet{Fivet2006}. The multiconfiguration model includes the following configurations: 5d$^{10}$, 5d$^9$\{5f, 6s, 6p, 6d, 6f, 7s, 7p, 7d, 8s\}, 5d$^8$\{6s$^2$, 6s5f, 6s6p, 6s6d, 6s6f, 6s7s, 6s7p, 6p$^2$, 7s$^2$, 7p$^2$\}, and 5d$^7$\{6s$^2$5f, 6s$^2$6p, 6s$^2$6d, 6s$^2$6f, 6s6p$^2$\}.
As in the cases of the previous ions considered in this work, the missing contribution from core--valence correlation in the HFR approach was treated by including a core--polarization potential with a \ion{Tl}{vii} ionic core, a dipole polarizability $\alpha_d = 3.25$ $a_0^3$, extrapolated from \citet{Fraga1976}, and a cut-off radius $r_c = 1.25$ $a_0$. The least-squares adjustment used 38 experimental levels from \citet{Joshi1990} and \citet{Wyart1992_TlIV_PbV_BiVI}, listed in Table~\ref{tab:tl-fit-levels}. The mean absolute relative deviation is reduced from 0.98\% to 0.068\% after the fit. The largest ab-initio discrepancies occur for the low-lying levels from 5d$^9$6s, where the initial errors are about 3.5--4.2\%; these are reduced to the 0.08--0.14\% range by the fit. Most 5d$^9$6d and 5d$^9$7s levels are reproduced within 0.01\% or better. The 5d$^9$6p odd-parity configuration is also substantially improved overall, although two levels that were already very close to experiment before adjustment show a slightly larger residual afterward. The strong reduction of the global deviation nevertheless demonstrates that the fitted Hamiltonian provides a markedly better representation of the observed energy spectrum and, consequently, of the configuration mixing used in the final transition-probability calculations.

\subsection{MCDHF models}
For each element considered in this work, a multireference (MR) was generated for each parity. Each MR corresponds to a set of spectroscopic configurations of the same parity. All calculations were performed using the EOL scheme for the states generated by the MR. Starting orbitals for the MR were obtained by performing a DHF calculation on the ground configuration of each element. A layer-by-layer approach was used to optimize the wavefunctions. The CSF expansion was generated by applying single and double (SD) excitations of electrons from specific subshells to an active set (AS) of correlation orbitals denoted by \{$n_1$s, $n_2$p, $n_3$d, $n_4$f, $n_5$g\}, where $n_i$ denotes the maximum principal quantum number $n$ associated with an azimuthal quantum number $l$. At each step, the previous orbitals were kept frozen, while only the newly added correlation orbitals were optimized. Correlation layers were added until convergence of the energy levels was observed. In order to refine the wavefunctions, further correlation effects were explored within a RCI approach, employing the optimized orbitals obtained from the final SD-MR expansion model. The effects of CSFs accounting for Core-Valence (CV) and Core-Core (CC) correlations were investigated using several RCI strategies. Details of the models for each ion are described below.

\textbf{\ion{As}{iii}}. The MR was constructed to describe the 21 experimentally known levels that arise from the odd-parity configurations 4s$^{2}$\{4p, 5p, 4f\} and 4p$^3$, and the even-parity configurations 4s$^{2}$\{5s, 6s, 4d, 5d, 5g\} and 4s4p$^2$ in \ion{As}{iii}. The orbital optimization was then carried out using a layer-by-layer approach, considering SD excitations from the valence subshells up to the last orbital set \{10s, 9p, 9d, 8f, 8g\}. Convergence of the computed energy levels was obtained after four correlation layers, with a mean absolute deviation of 2.8\% from the experimental energy levels. The correlation model was subsequently refined through an RCI calculation including single and restricted double excitations (SrD) from core orbitals, where restricted means that only one substitution is allowed in the orbital of interest. This extension of the CSF basis accounts for CV correlations involving the 3p and 3d subshells. This final MCDHF-RCI model led to a slight deterioration of the calculated energy levels compared with those obtained after the optimization step, producing a mean absolute deviation of 3.4\%. Nevertheless, it was adopted for the computation of the radiative parameters because it includes core-correlation effects, similarly to the CPOL correction considered in our HFR model.

\textbf{\ion{Se}{iii}}. No calculations were made for the \ion{Se}{iii} ion since accurate MCDHF transition data were recently published in the literature by \citet{Kitoviene2024_SeIIIBrIVKrV}. Levels arising from the experimentally known configurations 4s$^2$4p\{7s, 8s, 6d, 7d\} were not included in their model and therefore were not resolved compared to our HFR model. Their best model is similar to our \ion{As}{III} model, with CV correlations involving the 3s, 3p and 3d shells, achieving a relative agreement of 0.8\% with the experimental energy levels from \citet{Tauheed2012}.

\textbf{\ion{Hf}{iv}}. MCDHF calculations for \ion{Hf}{iv} were performed for the 14 targeted states belonging to the 4f$^{14}$\{6s, 7s, 8s, 5d, 6d\} configurations of even parity and the 4f$^{14}$\{6p, 7p, 5f, 6f\} configurations of odd parity. All these configurations were included in the MR. The orbital optimization step was performed using a layer-by-layer approach, taking into account SD excitations from the 4f and valence orbitals. The CSF expansion was generated up to the last active set \{12s, 11p, 11d, 8f, 7g\}. The mean absolute deviation obtained during the optimization step decreased from 2.5\% to 2.2\% upon inclusion of the last correlation layer. Finally, single and restricted double excitations (SrD) from the 5p core subshell were taken into account in an RCI calculation. Opening this subshell allowed us to further expand the CSF basis by including not only CV correlations but also CC correlations involving the 4f and 5p orbitals. Although this final MCDHF-RCI model slightly degraded the agreement with the experimental energy levels, yielding a mean absolute deviation of 3.1\%, it was retained for the computation of the radiative parameters for the same reason as evoked for \ion{As}{iii}.

\textbf{\ion{Tl}{iv}}. To model \ion{Tl}{iv}, the reference space was defined by the 5d$^{10}$, 5d$^9$\{6s, 7s, 6d, 7d\} even-parity configurations and the 5d$^9$6p odd-parity configuration, corresponding to the 38 experimentally known levels considered in this work. Valence-Valence (VV) correlations, generated using SD substitutions from the 5d, 6d, 7d, 6s, and 7s valence orbitals, were used for the orbital optimization procedure. The last correlation layer yielded a mean absolute deviation of 3.4\%, which was improved to 2.5\% in a further MCDHF-RCI model. This model includes, in addition to VV correlations, CSFs accounting for CV effects from the inner 4f orbital and was used for the computation of the transition parameters.\

\section{Results and discussion}
In the present work, the HFR calculations, which benefit from a semi-empirical adjustment of radial parameters to experimental energy levels, constitute the primary dataset intended for astrophysical applications. The independent MCDHF calculations, based on a relativistic treatment in which electron correlation is introduced explicitly through progressively enlarged configuration spaces, are used as a complementary benchmark to assess the reliability of the resulting radiative parameters. The comparisons between the HFR and MCDHF energy levels are given in Appendix \ref{app:levels}, while the oscillator strengths obtained by both methods are provided in Appendix \ref{app:transitions}. We note that, for Se III, no new MCDHF calculations were performed in this work, since recent accurate MCDHF data are already available in the literature and can be used for this independent comparison.

For the transition comparison, particular attention must be paid to internal reliability indicators provided by the two methods. In the HFR framework, a small cancellation factor (CF) indicates that the transition amplitude results from strong cancellation between different contributions and that the corresponding oscillator strength can therefore be very sensitive to relatively small changes in the wavefunction \citep{Cowan1981}. In the MCDHF calculations, the gauge disagreement $dT$ measures the consistency between the Babushkin and Coulomb forms of the transition probability and provides an internal estimate of the stability of the radiative data. Conversely, transitions with very small $\mathrm{CF}$ values or large $dT$ values should be treated with caution, as these indicators point to increased sensitivity to cancellation effects or gauge dependence. In the forthcoming analyses, the comparisons will be carried out for the complete set of radiative transitions of each ion, as well as for a restricted set including only strong transitions ($\log(gf)_{\mathrm{HFR}}\geq -1$) that satisfy the following criteria: $\mathrm{CF}\geq 0.05$ and $dT\leq 0.20$.

\begin{figure}[h]
\centering
\begin{minipage}[t]{0.49\textwidth}
    \centering
    \includegraphics[width=\linewidth]{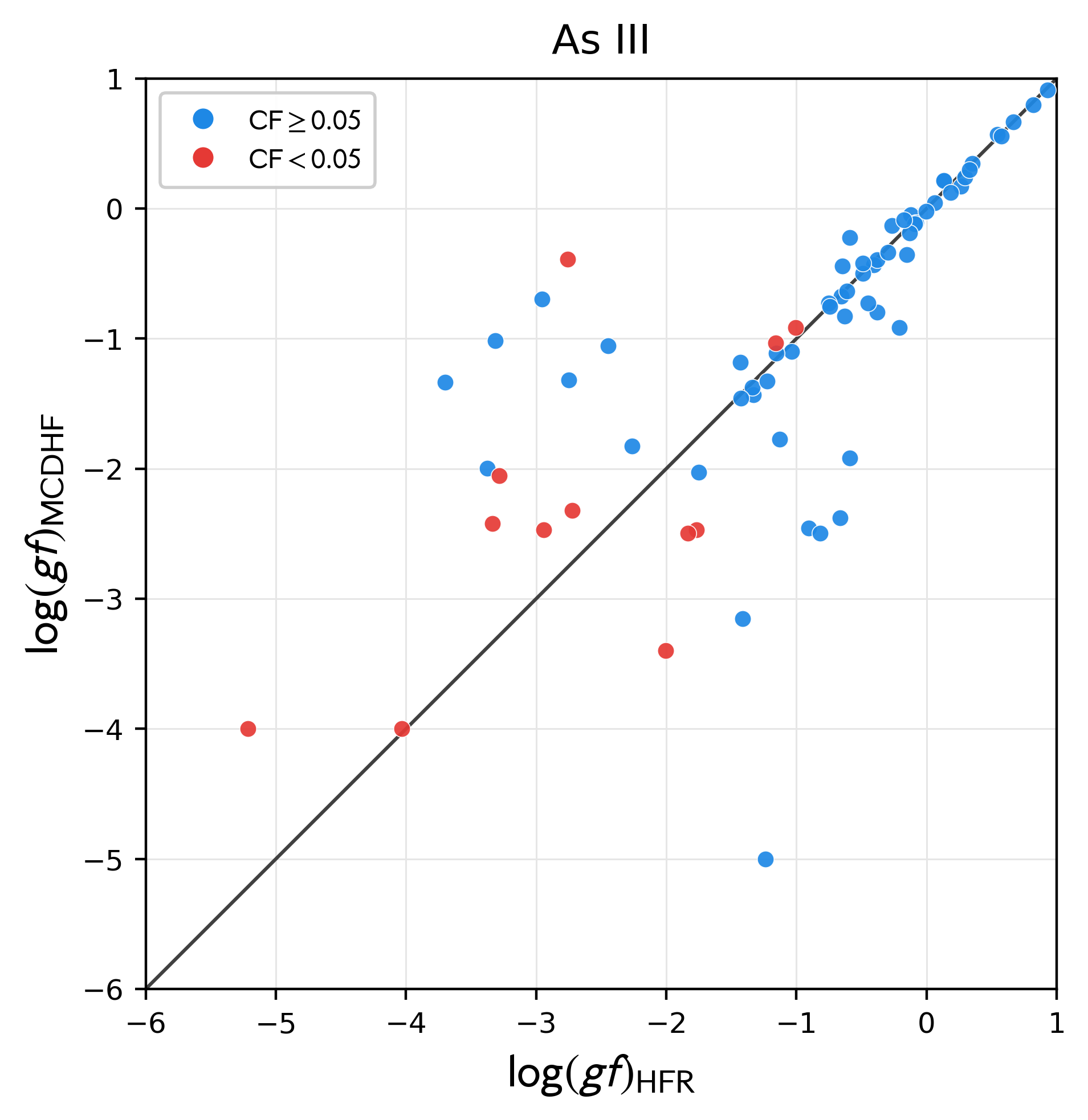}
    \textbf{(a)}
\end{minipage}
\hfill
\begin{minipage}[t]{0.49\textwidth}
    \centering
    \includegraphics[width=\linewidth]{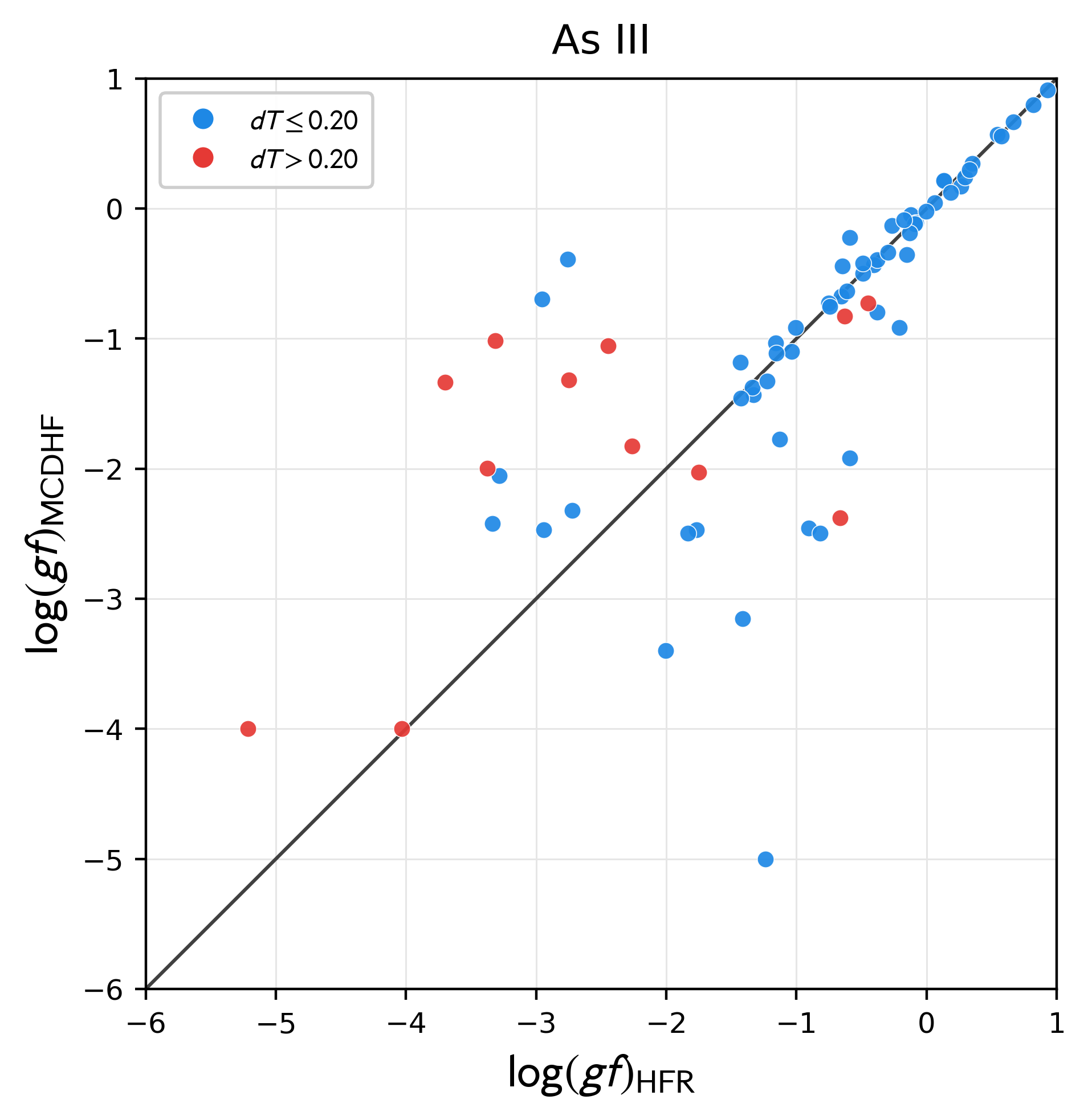}
    \textbf{(b)}
\end{minipage}
\caption{Comparison between HFR and MCDHF oscillator strengths for \ion{As}{iii}. 
(a) Transitions colour-coded according to the value of the HFR cancellation factor CF. 
(b) Transitions colour-coded according to the MCDHF gauge disagreement $dT$. 
The solid diagonal line indicates perfect agreement between the two calculations.}
\label{fig:as-gf-comparison}
\end{figure}

\begin{figure}[h]
\centering
\begin{minipage}[t]{0.49\textwidth}
    \centering
    \includegraphics[width=\linewidth]{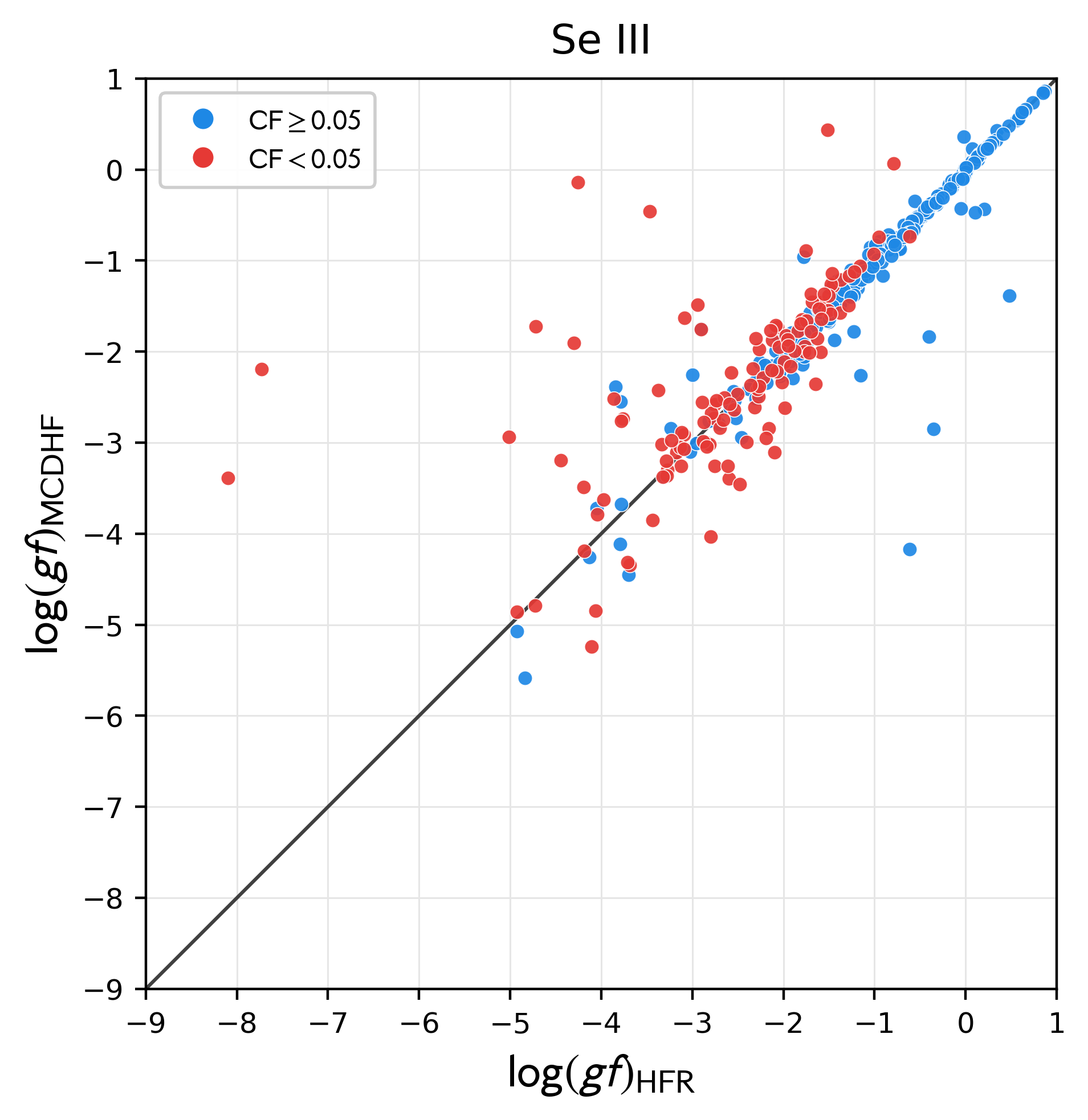}
    \textbf{(a)}
\end{minipage}
\hfill
\begin{minipage}[t]{0.49\textwidth}
    \centering
    \includegraphics[width=\linewidth]{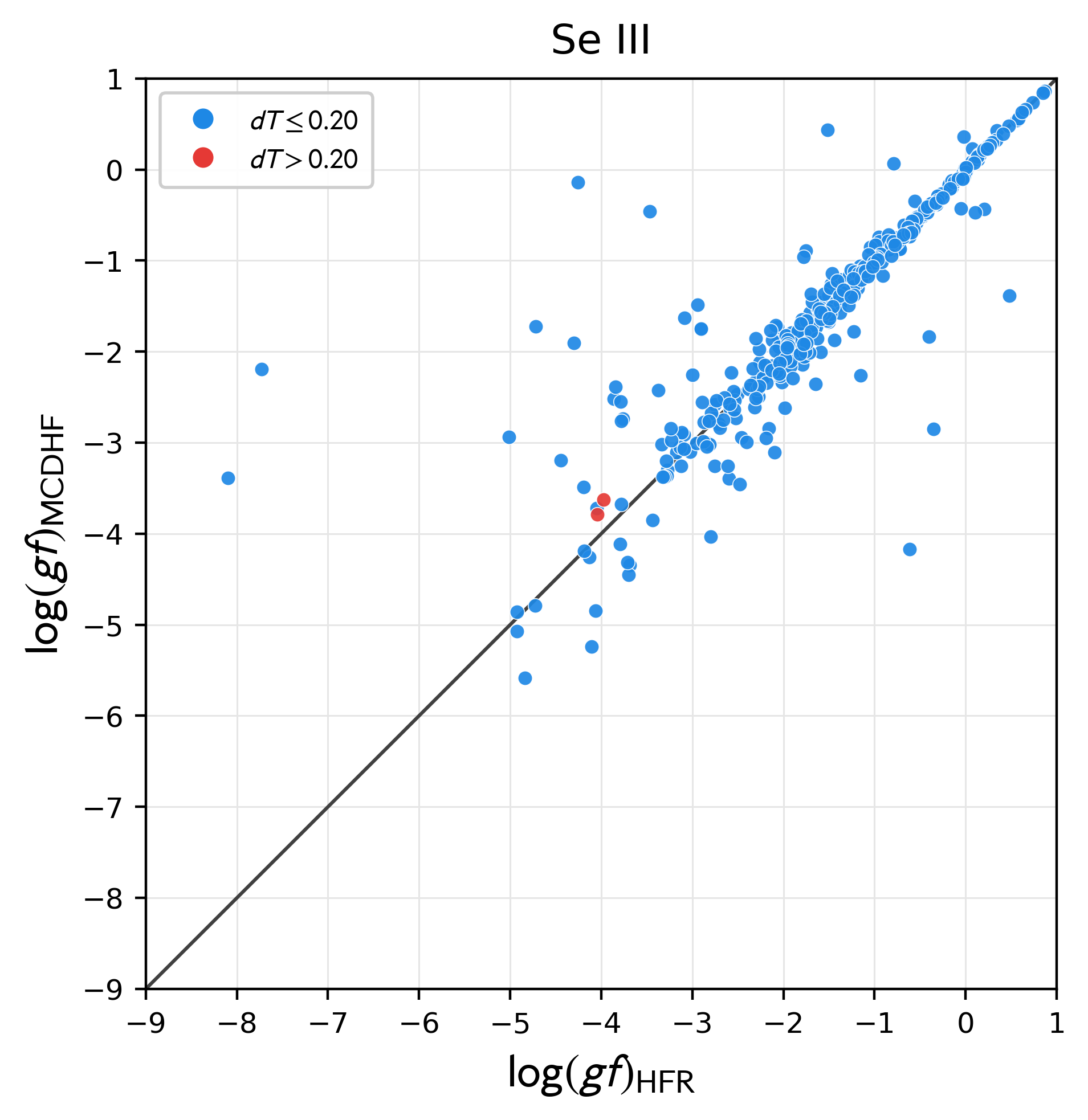}
    \textbf{(b)}
\end{minipage}
\caption{Comparison between HFR and MCDHF oscillator strengths for \ion{Se}{iii}. 
(a) Transitions colour-coded according to the value of the HFR cancellation factor CF. 
(b) Transitions colour-coded according to the MCDHF gauge disagreement $dT$. 
The solid diagonal line indicates perfect agreement between the two calculations.}
\label{fig:se-gf-comparison}
\end{figure}

\begin{figure}[H]
\centering
\begin{minipage}[t]{0.49\textwidth}
    \centering
    \includegraphics[width=\linewidth]{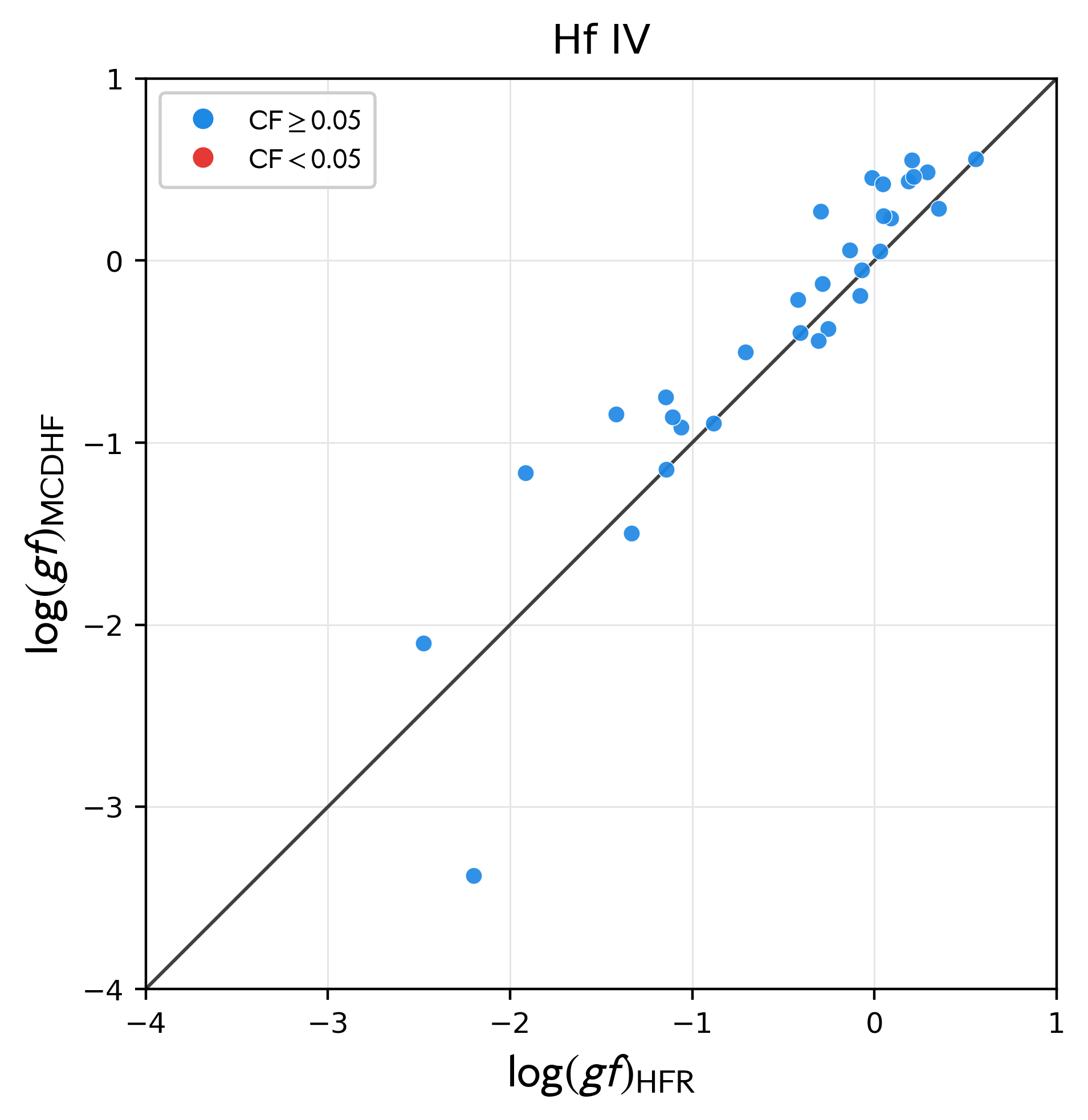}
    \textbf{(a)}
\end{minipage}
\hfill
\begin{minipage}[t]{0.49\textwidth}
    \centering
    \includegraphics[width=\linewidth]{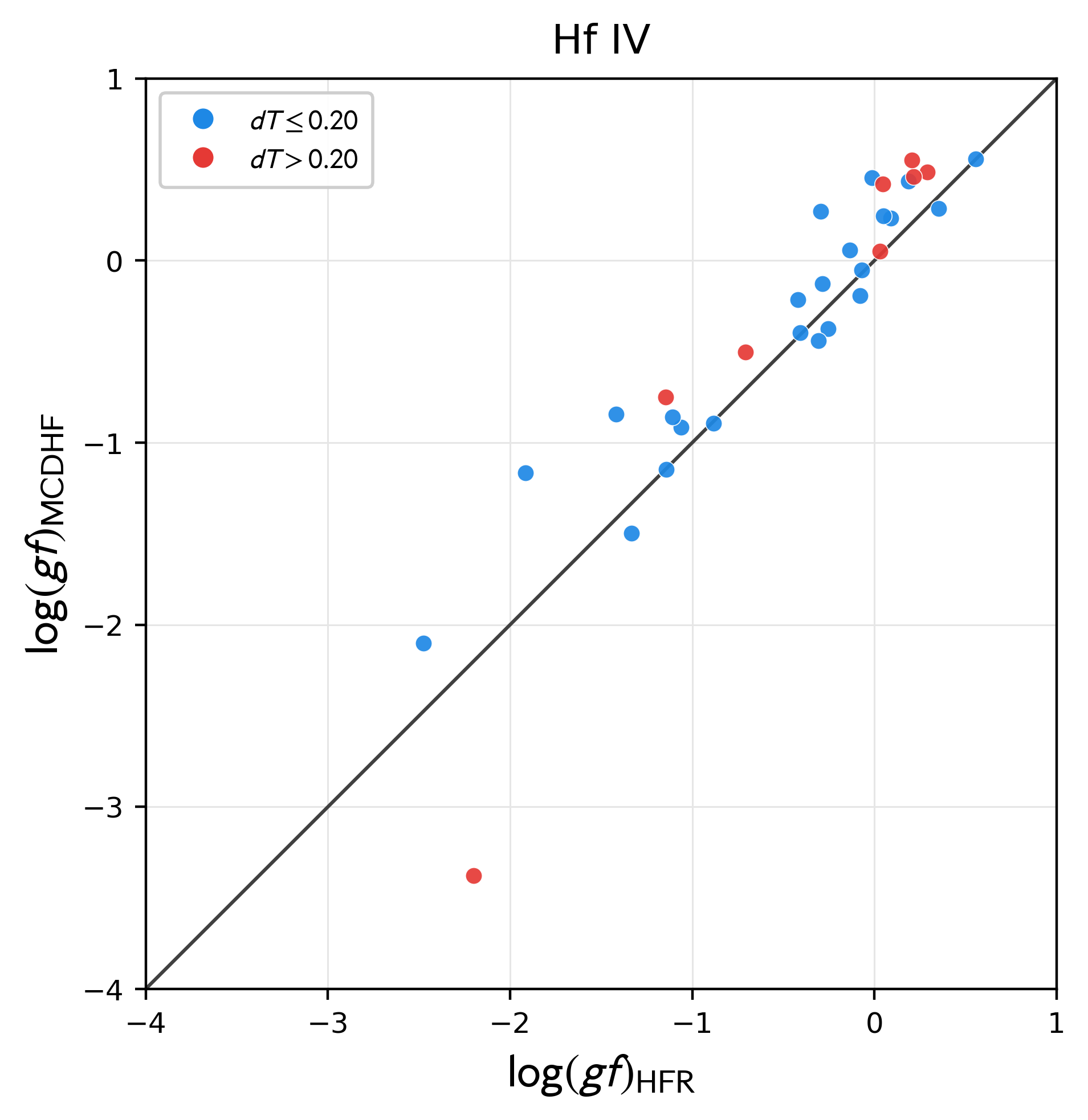}
    \textbf{(b)}
\end{minipage}
\caption{Comparison between HFR and MCDHF oscillator strengths for \ion{Hf}{iv}. 
(a) Transitions colour-coded according to the value of the HFR cancellation factor CF. 
(b) Transitions colour-coded according to the MCDHF gauge disagreement $dT$. 
The solid diagonal line indicates perfect agreement between the two calculations.}
\label{fig:hf-gf-comparison}
\end{figure}

\begin{figure}[H]
\centering
\begin{minipage}[t]{0.49\textwidth}
    \centering
    \includegraphics[width=\linewidth]{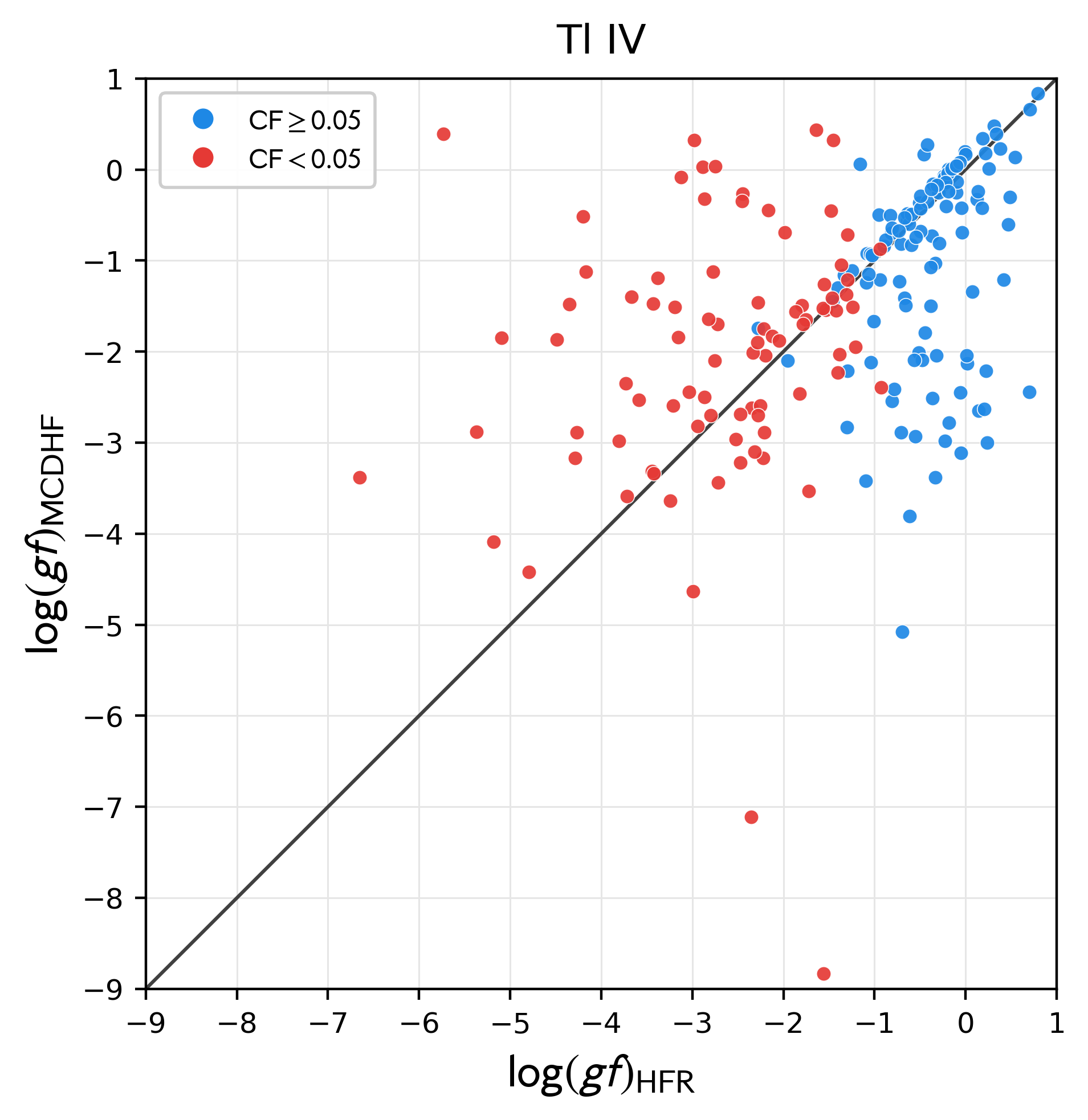}
    \textbf{(a)}
\end{minipage}
\hfill
\begin{minipage}[t]{0.49\textwidth}
    \centering
    \includegraphics[width=\linewidth]{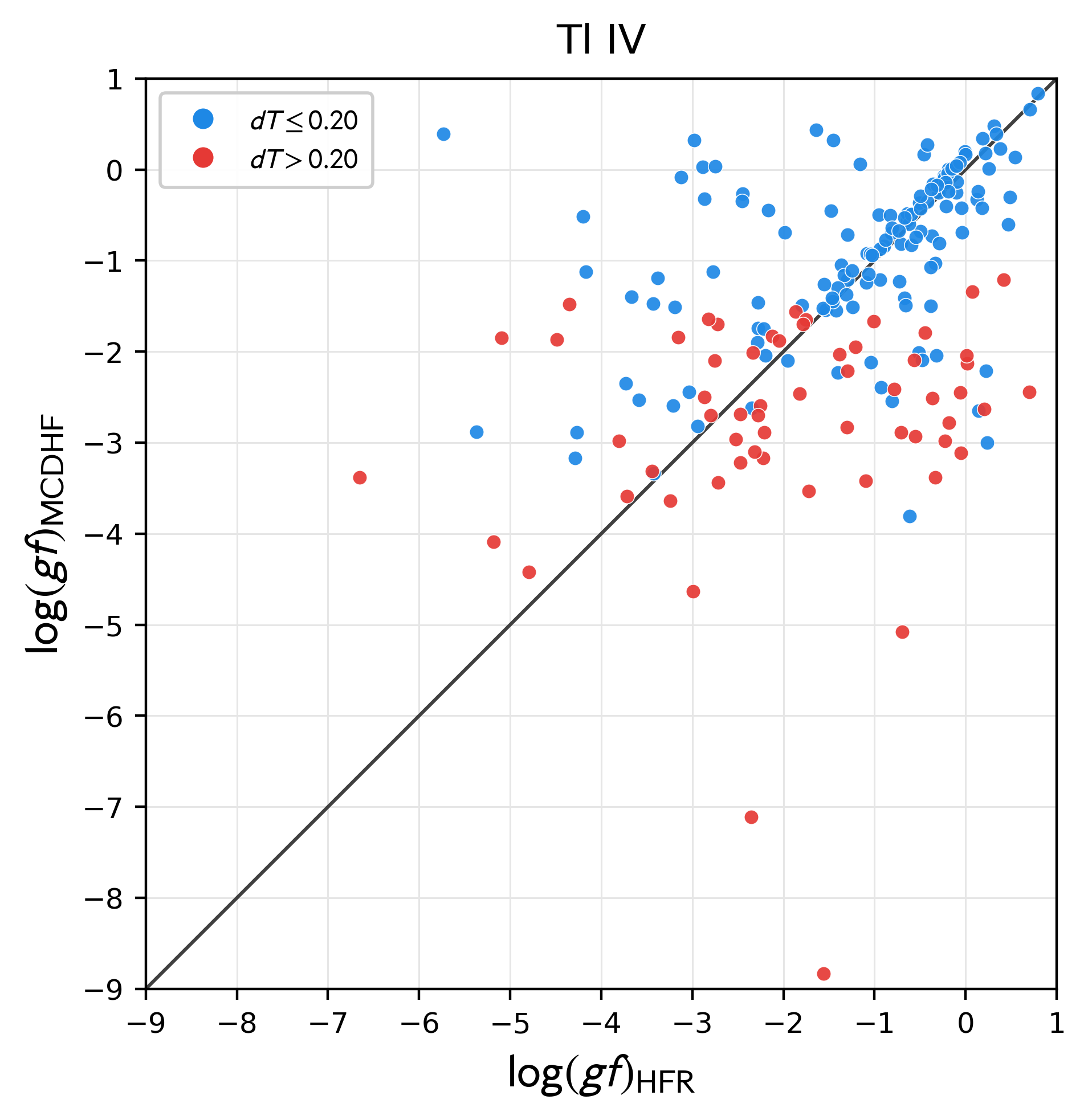}
    \textbf{(b)}
\end{minipage}
\caption{Comparison between HFR and MCDHF oscillator strengths for \ion{Tl}{iv}. 
(a) Transitions colour-coded according to the value of the HFR cancellation factor CF. 
(b) Transitions colour-coded according to the MCDHF gauge disagreement $dT$. 
The solid diagonal line indicates perfect agreement between the two calculations.}
\label{fig:tl-gf-comparison}
\end{figure}

To quantify the agreement between the two calculations, we use the mean absolute relative difference in oscillator strength,
\begin{equation}
	\Delta_{gf}=\frac{\left|gf_{\rm HFR}-gf_{\rm MCDHF}\right|}
	{\max\left(gf_{\rm HFR},gf_{\rm MCDHF}\right)}\,.
\end{equation}
The values discussed below use the MCDHF oscillator strengths in the Babushkin gauge, consistent with the MCDHF values retained in the transition tables for \ion{As}{III} (Table \ref{tab:as-hfrcpol-transitions}), \ion{Hf}{IV} (Table \ref{tab:hf-hfrcpol-transitions}), and \ion{Tl}{IV} (Table \ref{tab:tl-hfrcpol-transitions}); for \ion{Se}{III}, the corresponding values from the published MCDHF dataset \citep{Kitoviene2024_SeIIIBrIVKrV} are used. The comparison is first made over all transitions common to the two approaches and then repeated for the restricted subset defined by the above-mentioned criteria on CF and $dT$.
The comparison between the HFR and MCDHF $\log(gf)$ is shown for \ion{As}{iii}, \ion{Se}{iii}, \ion{Hf}{iv} and \ion{Tl}{iv} in Figure \ref{fig:as-gf-comparison}, Figure \ref{fig:se-gf-comparison}, Figure \ref{fig:hf-gf-comparison} and Figure \ref{fig:tl-gf-comparison}, respectively. 

\subsection{\ion{As}{iii}}

The comparison of the \ion{As}{iii} oscillator strengths showed that weak transitions account for most of the dispersion between HFR and MCDHF. For the complete set of transitions in the selected MCDHF correlation model, the mean absolute relative difference in $gf$ is about 41\%. The scatter is however dominated by weak transitions particularly sensitive to the details of the atomic wavefunctions. Applying the above-mentioned reliability criteria for the strong transitions ($\log(gf)_{\mathrm{HFR}}\geq -1$, $\mathrm{CF}\geq 0.05$ and $dT \leq 0.20$) reduces the mean scatter to about 20\%. The strong transitions therefore not affected by cancellation effects or gauge disagreement show a substantially better agreement than the complete transition set, showing that a substantial part of the dispersion in the complete dataset originates from transitions that are either weak or internally more sensitive in one of the two calculations.

The astrophysical relevance of these calculations has already been demonstrated in the parallel analysis of the heavy-metal subdwarf \lsiv\ by \citet{Dorsch2026}. Five \ion{As}{iii} lines calculated in the present work were used in the abovementioned analysis, i.e. the lines at 1172.150, 1209.280, 1274.310, 3922.498, and 4037.035~\AA. The new oscillator strengths made it possible to treat the ultraviolet and optical \ion{As}{iii} lines consistently in the spectral synthesis and derive the As III abundance in \lsiv. Several additional strong transitions in the present calculations are attractive candidates for future searches. In particular, the transitions with Ritz wavelengths near 2926.97 and 2982.76~\AA\ have $\log(gf)_{\mathrm{HFR}}=0.331$ and 0.577, respectively; the corresponding MCDHF oscillator strengths differ from the HFR values by only about 7\% and 4\%. Strong lines are also predicted near 4032.26 and 4033.56~\AA, including transitions with $\log(gf)$ close to 0.9 and 0.8, for which the HFR--MCDHF differences in $gf$ are about 5\%. Their favorable CF and $dT$ values make them particularly useful candidates under suitable conditions of wavelength coverage, level populations, and line blending.

\subsection{\ion{Se}{iii}}
The comparison for \ion{Se}{iii} is performed only for transitions common to the present HFR calculations and the published independent MCDHF dataset \citep{Kitoviene2024_SeIIIBrIVKrV}. This yields 382 transitions, fewer than the total number of HFR transitions because the two calculations do not cover exactly the same configuration sets. The detailed analysis performed for \ion{Se}{iii} found a mean absolute relative difference of approximately 27\% between the HFR and MCDHF oscillator strengths for the complete set of transitions. When only strong transitions ($\log(gf)_{\mathrm{HFR}}\geq -1$) with $\mathrm{CF}\geq 0.05$ and $dT\leq 0.20$ are retained, the mean difference decreased to about 12\%. 
The reduction in the scatter after applying these criteria is clearly associated with the removal of weak transitions (with limited or no interest for astrophysical application) affected by cancellation effects or a gauge disagreement. For the strong lines, the two independent calculations therefore give oscillator strengths that are generally consistent.

The new \ion{Se}{iii} data were directly used in the abundance analysis of \lsiv\ presented by \citet{Dorsch2026}. Table~1 of that work contains 19 \ion{Se}{iii} spectral lines based on the present HFR calculations, predominantly in the optical UVES spectrum, and these lines provide the main constraint on the selenium abundance. In the ultraviolet, the \ion{Se}{iii} 1245.988~\AA\ line is securely identified.

\subsection{\ion{Hf}{iv}}

For \ion{Hf}{iv}, the MCDHF calculation targeted a more restricted set of states than the HFR model. The experimentally identified levels assigned to the 4f$^{13}$5d$^2$ configuration were not retained in the final MCDHF multireference. Exploratory calculations showed that these states were strongly mixed with 4f$^{13}$5d6s, with the 4f$^{13}$5d$^2$ component contributing only about 10--20\% to the corresponding wavefunctions, while 4f$^{13}$5d6s was generally dominant, preventing an unambiguous identification of the corresponding MCDHF levels. They were therefore excluded from the final MCDHF model. Consequently, only transitions connecting levels represented and reliably identified in both atomic models can be compared, yielding 31 common radiative transitions despite the substantially larger transition set available from the HFR calculations. Among the radiative transitions common to the two calculations, the mean relative difference in $gf$ is about 37\%, but this value is influenced by weak transitions. For the strong transitions satisfying the abovementioned criteria, the agreement improves as the mean absolute difference is approximately 31\%.
A few \ion{Hf}{iv} transitions present sizeable differences in their HFR and MCDHF oscillator strengths, even though neither CF nor $dT$ indicates an obvious numerical instability. For example, the transitions near 647.39 and 665.65~\AA\ have HFR $\log(gf)$ values of about $-0.29$ and $-0.01$, respectively, whereas the corresponding MCDHF values are about $0.27$ and $0.46$. The resulting differences in $gf$ are approximately 73\% and 66\%, despite favorable CF and $dT$ values. Conversely, several other strong \ion{Hf}{iv} transitions show much closer agreement, so the remaining dispersion cannot be attributed to a systematic offset between the two methods.

The \ion{Hf}{iv} calculations were developed specifically to support the analysis of \EC. Six particularly strong \ion{Hf}{iv} lines were identified in the HST spectrum by \citet{Dorsch2026}, at 1305.241, 1357.399, 1390.390, 1491.670, 1528.820, and 1717.210~\AA. Several additional transitions in the present line list are promising for future searches. Two examples are the lines with Ritz wavelengths near 1366.359 and 1572.030~\AA, for which the HFR/MCDHF values are $-0.308/-0.439$ and $-0.071/-0.052$, respectively. In terms of oscillator strengths, the corresponding differences are approximately 26\% and 4\%. Both transitions have large cancellation factors ($\mathrm{CF}\simeq 0.96$) and a small gauge disagreement ($dT<0.20$), making them particularly reliable according to the adopted criteria and to the good agreement between their HFR and MCDHF oscillator strengths. Their actual observability will, of course, also depend on the population of the lower level, the hafnium abundance, blending with other metal lines, and the available spectral coverage.

\subsection{\ion{Tl}{iv}}

\ion{Tl}{iv} displays the largest dispersion between the two calculations when the complete transition set is considered. For the 193 transitions common to HFR and MCDHF, the mean $\Delta_{gf}$ is approximately 62\%. A substantial fraction of this scatter is associated with weak transitions affected by strong cancellation effects in the HFR calculation. In some particularly striking cases, HFR predicts an extremely small oscillator strength together with a very small CF, whereas MCDHF predicts a substantially stronger transition while maintaining a small gauge disagreement. One example is the transition near 788.88~\AA, for which $\log(gf)_{\rm HFR}\simeq-4.20$ with a CF close to zero, whereas the MCDHF value is about $-0.52$ with $dT\simeq0.02$. In such a case, the larger disagreement between the two methods should not by itself be interpreted as evidence against the MCDHF value, since the HFR result is simultaneously affected by significant cancellation effects.

Applying the restrictions defined earlier to the strong transitions reduces the sample to 61 transitions and the mean difference to approximately 40\%. The improvement from 62\% to 40\% confirms that weak and internally sensitive transitions contribute strongly to the global dispersion. Although the agreement is less tight than for \ion{As}{iii}, \ion{Se}{iii} and Hf IV, the two independent approaches therefore generally predict the oscillator strengths of the strong and internally stable \ion{Tl}{iv} transitions within less than 40\% on average. 

The importance of these data is demonstrated directly by the HST analysis of \EC\ in \citet{Dorsch2026}. Table~1 of that work lists 18 \ion{Tl}{iv} spectral components calculated using the present atomic data. Because several transitions exhibit resolved hyperfine splitting, these 18 line entries correspond to 10 distinct electronic transitions. The strongest observed features include the hyperfine doublets near 1272.9, 1337.1, 1377.7, and 1404.6~\AA, together with the single line at 1374.619~\AA. The availability of oscillator strengths for these transitions made a quantitative thallium abundance determination possible.

The present calculations also contain several strong and theoretically stable \ion{Tl}{iv} transitions that are not among the lines listed as identified in that analysis. Particularly interesting examples occur at Ritz wavelengths of 1171.506, 1181.641, 1221.235, and 1243.667~\AA. Their HFR/MCDHF $\log(gf)$ values are $-0.621/-0.600$, $-0.595/-0.491$, $-0.496/-0.426$, and $-0.189/-0.242$, respectively, corresponding to absolute differences in $gf$ of about 5\%, 27\%, 17\%, and 11\%. All four satisfy the adopted CF and $dT$ reliability criteria. They therefore represent favorable candidates for searches in spectra of other heavy-metal hot subdwarfs, or in higher-quality spectra of \EC, provided that the corresponding lower levels are sufficiently populated and that blending does not prevent their detection.

\subsection{Summary}
Overall, the oscillator strengths obtained by both methods are found to be in good agreement when restricting  to strong transitions that are not associated with very small CF (strong cancellation effects) or large $dT$ (significant gauge disagreement). The mean differences between the HFR and the MCDHF oscillator strengths for the restricted set of strong transitions are always lower than 40\% for the four ions considered in this work (20\% for \ion{As}{iii}, 12\% for \ion{Se}{iii}, 31\% for \ion{Hf}{iv}, and 40\% for \ion{Tl}{iv}). The comparison therefore supports the overall consistency of the HFR oscillator strengths for the strong and most relevant transitions to stellar spectroscopy while also identifying individual transitions for which a larger theoretical uncertainty should be kept in mind.

The astrophysical usefulness of these data has already been demonstrated in the companion analysis by \citet{Dorsch2026}, in which the HFR oscillator strengths were used to model 48 observed spectral lines from the four ions and to derive their abundances in \lsiv\ and \EC. As discussed above for each ion, the present line lists also contain several additional strong transitions with favorable internal reliability indicators and good agreement between the two calculations, providing promising targets for future stellar spectroscopic analyses. Their detectability will ultimately depend on level populations, elemental abundances, spectral coverage, and line blending.


\section{Conclusions}

In this work, we have presented new radiative data for \ion{As}{iii}, \ion{Se}{iii}, \ion{Hf}{iv}, and \ion{Tl}{iv}, with the aim of improving the atomic data available for the quantitative analysis of heavy-metal hot-subdwarf spectra. The HFR calculations, including core-polarization effect corrections and a semi-empirical adjustment of radial parameters to experimental energy levels, provide the main set of oscillator strengths intended for astrophysical applications. Their reliability was assessed through comparison with new independent MCDHF calculations or, in the case of \ion{Se}{iii}, with recent MCDHF data available in the literature. The comparison shows that the largest discrepancies between the two approaches are mainly associated with weak transitions and with transitions affected by strong cancellation effects in HFR or significant gauge disagreement in MCDHF. When the analysis is restricted to stronger and reliable transitions, the agreement improves substantially, with mean differences of about 20\%, 12\%, 31\%, and 40\% for \ion{As}{iii}, \ion{Se}{iii}, \ion{Hf}{iv}, and \ion{Tl}{iv}, respectively. Although agreement between two theoretical approaches cannot by itself establish the accuracy of an individual transition, the consistency obtained for the strongest lines provides an important independent assessment of the radiative data most relevant to stellar spectroscopy.

The astrophysical relevance of the calculations has already been demonstrated through their application to the heavy-metal hot subdwarfs \lsiv\ and \EC\ in the companion spectroscopic analysis of \citet{Dorsch2026}. In total, oscillator strengths from the present calculations were used to model 48 observed spectral lines belonging to the four ions, enabling quantitative abundance determinations of As and Se in \lsiv\ and of Hf and Tl in \EC. This provides a direct practical validation of the usefulness of the calculated data in stellar-atmosphere modeling and extends the set of heavy elements for which quantitative abundance analyses can be performed in this temperature regime. The comparison between independent atomic-structure methods also provides useful guidance for identifying transitions for which larger theoretical uncertainties should be considered in future spectroscopic applications.

Beyond the lines already identified in these stars, the calculated transition lists contain several additional strong and reliable transitions and, in many cases, close agreement between the HFR and MCDHF oscillator strengths. These transitions provide promising targets for future searches in heavy-metal hot subdwarfs and other chemically peculiar stars in which ionization stages III--IV are expected to dominate. Further progress will benefit from improved experimental level energies and laboratory spectroscopy, particularly for transitions whose predicted wavelengths remain uncertain, as well as from new high-resolution ultraviolet and optical observations. More generally, extending this combined strategy to additional poorly studied heavy ions will help close the remaining gaps in atomic data that currently limit the interpretation of chemically peculiar stellar spectra and the use of heavy-element abundances as diagnostics of diffusion and nucleosynthesis processes.


\vspace{6pt} 


\authorcontributions{All the authors contribute equally to this paper.}


\dataavailability{The data can be made available upon reasonable request.}

\acknowledgments{
MD was supported by the Deutsches Zentrum für Luft- und Raumfahrt (DLR) through grant 50-OR-2304. JD and LM are supported by the FWO and F.R.S.-FNRS under the Excellence of Science (EOS) programme (numbers O.0004.22 and O022818F). PQ is F.R.S.-FNRS Research Director.}

\appendix
\renewcommand{\thetable}{\thesection\arabic{table}}
\makeatletter
\@addtoreset{table}{section}
\makeatother
\section{Semi-empirical HFR fitting data}
\label{app:hfrfit}
This appendix collects the HFR and HFR+fit radial parameters used in Section~3.1. 

\begingroup
\scriptsize
\setlength{\tabcolsep}{3.5pt}
\begin{longtable}{@{}llrrr@{}}
\caption{HFR and HFR+fit parameters for \ion{As}{iii}.}\label{tab:as-fit-params}\\
\toprule
Configuration & Parameter & HFR (cm$^{-1}$) & HFR+fit (cm$^{-1}$) & Ratio \\
\midrule
\endfirsthead
\multicolumn{5}{c}{\tablename~\thetable\ (continued)}\\
\toprule
Configuration & Parameter & HFR (cm$^{-1}$) & HFR+fit (cm$^{-1}$) & Ratio \\
\midrule
\endhead
\midrule \multicolumn{5}{r}{Continued on next page}\\ \endfoot
\bottomrule \endlastfoot
 4s$^{2}$4p & $E_{\rm av}$ & 8\,705.9 & 9\,599.2 & -- \\
 & $\zeta(4p)$ & 1\,806.3 & 2\,031.1 & 1.124 \\
 4s$^{2}$5p & $E_{\rm av}$ & 139\,980.1 & 133\,340.3 & -- \\
 & $\zeta(5p)$ & 437.6 & 497.9 & 1.138 \\
 4s$^{2}$4f & $E_{\rm av}$ & 171\,916.5 & 171\,200.2 & -- \\
 & $\zeta(4f)$ & 0.5 & 0.4 & 0.800 \\
 4p$^{3}$ & $E_{\rm av}$ & 190\,799.5 & 158\,506.0 & -- \\
 & $\zeta(4p)$ & 1\,825.2 & 0.0 & 0.000 \\
 & $\alpha(4p)$ & 0.0 & -111.1 & -- \\
 4s4p$^{2}$ & $E_{\rm av}$ & 89\,502.2 & 90\,191.0 & -- \\
 & $\zeta(4p)$ & 1\,813.2 & 2\,188.2 & 1.207 \\
 & $\alpha(4p)$ & 0.0 & 159.6 & -- \\
 & $F^2(4p,4p)$ & 49\,052.6 & 46\,330.4 & 0.945 \\
 & $G^1(4s,4p)$ & 66\,178.8 & 56\,613.2 & 0.855 \\
 4s$^{2}$5s & $E_{\rm av}$ & 114\,819.8 & 113\,916.0 & -- \\
 4s$^{2}$6s & $E_{\rm av}$ & 170\,776.3 & 163\,291.2 & -- \\
 4s$^{2}$4d & $E_{\rm av}$ & 119\,684.6 & 118\,762.3 & -- \\
 & $\zeta(4d)$ & 62.7 & 146.9 & 2.343 \\
 4s$^{2}$5d & $E_{\rm av}$ & 173\,097.9 & 165\,575.9 & -- \\
 & $\zeta(5d)$ & 25.6 & 34.8 & 1.359 \\
 4s$^{2}$5g & $E_{\rm av}$ & 196\,275.4 & 189\,077.3 & -- \\
\end{longtable}
\noindent\footnotesize The HFR column gives the ab initio parameter values and the HFR+fit column the fitted values. Ratio = (HFR+fit)/HFR. Ratios are not quoted for $E_{\rm av}$ or for effective parameters whose HFR value is zero.
\endgroup

\begingroup
\scriptsize
\setlength{\tabcolsep}{3.5pt}
\begin{longtable}{@{}llrrr@{}}
\caption{HFR and HFR+fit parameters for \ion{Se}{iii}.}\label{tab:se-fit-params}\\
\toprule
Configuration & Parameter & HFR (cm$^{-1}$) & HFR+fit (cm$^{-1}$) & Ratio \\
\midrule
\endfirsthead
\multicolumn{5}{c}{\tablename~\thetable\ (continued)}\\
\toprule
Configuration & Parameter & HFR (cm$^{-1}$) & HFR+fit (cm$^{-1}$) & Ratio \\
\midrule
\endhead
\midrule \multicolumn{5}{r}{Continued on next page}\\ \endfoot
\bottomrule \endlastfoot
 4s4p$^{3}$ & $E_{\rm av}$ & 115\,423.4 & 115\,815.7 & -- \\
 & $\zeta(4p)$ & 2\,371.6 & 2\,716.2 & 1.145 \\
 & $\alpha(4p)$ & 0.0 & -157.9 & -- \\
 & $F^2(4p,4p)$ & 52\,865.8 & 47\,172.6 & 0.892 \\
 & $G^1(4s,4p)$ & 71\,483.5 & 58\,114.4 & 0.813 \\
 4s$^{2}$4p5s & $E_{\rm av}$ & 138\,037.0 & 137\,624.3 & -- \\
 & $\zeta(4p)$ & 2\,621.4 & 2\,844.7 & 1.085 \\
 & $G^1(4p,5s)$ & 5\,622.8 & 4\,068.5 & 0.724 \\
 4s$^{2}$4p6s & $E_{\rm av}$ & 198\,325.3 & 190\,602.1 & -- \\
 & $\zeta(4p)$ & 2\,675.7 & 2\,880.8 & 1.077 \\
 & $G^1(4p,6s)$ & 1\,640.9 & 1\,183.8 & 0.721 \\
 4s$^{2}$4p7s & $E_{\rm av}$ & 223\,811.1 & 216\,057.5 & -- \\
 & $\zeta(4p)$ & 2\,689.7 & 2\,900.2 & 1.078 \\
 & $G^1(4p,7s)$ & 737.8 & 505.5 & 0.685 \\
 4s$^{2}$4p8s & $E_{\rm av}$ & 237\,162.5 & 229\,499.7 & -- \\
 & $\zeta(4p)$ & 2\,695.2 & 2\,907.3 & 1.079 \\
 & $G^1(4p,8s)$ & 400.6 & 268.4 & 0.670 \\
 4s$^{2}$4p4d & $E_{\rm av}$ & 141\,844.9 & 141\,218.5 & -- \\
 & $\zeta(4p)$ & 2\,543.3 & 2\,781.9 & 1.094 \\
 & $\zeta(4d)$ & 84.0 & 167.5 & 1.994 \\
 & $F^2(4p,4d)$ & 34\,072.9 & 29\,285.7 & 0.860 \\
 & $G^1(4p,4d)$ & 38\,150.2 & 31\,861.5 & 0.835 \\
 & $G^3(4p,4d)$ & 23\,265.3 & 19\,645.8 & 0.844 \\
 4s$^{2}$4p5d & $E_{\rm av}$ & 200\,500.6 & 192\,639.2 & -- \\
 & $\zeta(4p)$ & 2\,664.4 & 2\,872.8 & 1.078 \\
 & $\zeta(5d)$ & 31.0 & 42.4 & 1.368 \\
 & $F^2(4p,5d)$ & 9\,768.1 & 7\,946.2 & 0.813 \\
 & $G^1(4p,5d)$ & 8\,176.8 & 5\,592.6 & 0.684 \\
 & $G^3(4p,5d)$ & 5\,381.3 & 3\,385.8 & 0.629 \\
 4s$^{2}$4p6d & $E_{\rm av}$ & 224\,977.6 & 217\,124.2 & -- \\
 & $\zeta(4p)$ & 2\,685.3 & 2\,838.9 & 1.057 \\
 & $\zeta(6d)$ & 15.2 & 0.0 & 0.000 \\
 & $F^2(4p,6d)$ & 4\,285.2 & 3\,446.6 & 0.804 \\
 & $G^1(4p,6d)$ & 3\,346.9 & 2\,831.8 & 0.846 \\
 & $G^3(4p,6d)$ & 2\,265.1 & 1\,916.5 & 0.846 \\
 4s$^{2}$4p7d & $E_{\rm av}$ & 237\,843.7 & 230\,032.8 & -- \\
 & $\zeta(4p)$ & 2\,692.8 & 2\,814.8 & 1.045 \\
 & $\zeta(7d)$ & 8.7 & 0.0 & 0.000 \\
 & $F^2(4p,7d)$ & 2\,300.1 & 1\,368.5 & 0.595 \\
 & $G^1(4p,7d)$ & 1\,745.9 & 1\,477.2 & 0.846 \\
 & $G^3(4p,7d)$ & 1\,198.5 & 1\,014.0 & 0.846 \\
 4s4p$^{2}$5p & $E_{\rm av}$ & 261\,587.0 & 281\,586.5 & -- \\
 4s$^{2}$4p5g & $E_{\rm av}$ & 226\,284.9 & 246\,284.5 & -- \\
 4s$^{2}$4p$^{2}$ & $E_{\rm av}$ & 15\,622.9 & 15\,562.8 & -- \\
 & $\zeta(4p)$ & 2\,364.7 & 2\,617.2 & 1.107 \\
 & $\alpha(4p)$ & 0.0 & -11.7 & -- \\
 & $F^2(4p,4p)$ & 52\,823.1 & 41\,476.0 & 0.785 \\
 4s$^{2}$4p5p & $E_{\rm av}$ & 164\,710.7 & 157\,836.4 & -- \\
 & $\zeta(4p)$ & 2\,668.9 & 2\,818.4 & 1.056 \\
 & $\zeta(5p)$ & 494.9 & 549.4 & 1.110 \\
 & $F^2(4p,5p)$ & 14\,130.6 & 10\,136.2 & 0.717 \\
 & $G^0(4p,5p)$ & 3\,867.6 & 2\,475.0 & 0.640 \\
 & $G^2(4p,5p)$ & 4\,516.9 & 1\,389.3 & 0.308 \\
\end{longtable}
\noindent\footnotesize The HFR column gives the ab initio parameter values and the HFR+fit column the fitted values. Ratio = (HFR+fit)/HFR. Ratios are not quoted for $E_{\rm av}$ or for effective parameters whose HFR value is zero.
\endgroup

\begingroup
\scriptsize
\setlength{\tabcolsep}{3.5pt}
\begin{longtable}{@{}llrrr@{}}
\caption{HFR and HFR+fit parameters for \ion{Hf}{iv}.}\label{tab:hf-fit-params}\\
\toprule
Configuration & Parameter & HFR (cm$^{-1}$) & HFR+fit (cm$^{-1}$) & Ratio \\
\midrule
\endfirsthead
\multicolumn{5}{c}{\tablename~\thetable\ (continued)}\\
\toprule
Configuration & Parameter & HFR (cm$^{-1}$) & HFR+fit (cm$^{-1}$) & Ratio \\
\midrule
\endhead
\midrule \multicolumn{5}{r}{Continued on next page}\\ \endfoot
\bottomrule \endlastfoot
 4f$^{14}$6p & $E_{\rm av}$ & 73\,778.0 & 74\,053.3 & -- \\
 & $\zeta(6p)$ & 6\,092.8 & 6\,417.7 & 1.053 \\
 4f$^{14}$7p & $E_{\rm av}$ & 160\,838.6 & 161\,977.9 & -- \\
 & $\zeta(7p)$ & 2\,404.9 & 2\,233.5 & 0.929 \\
 4f$^{14}$5f & $E_{\rm av}$ & 154\,150.7 & 156\,945.8 & -- \\
 & $\zeta(5f)$ & 33.0 & 32.3 & 0.979 \\
 4f$^{14}$6f & $E_{\rm av}$ & 195\,399.4 & 196\,106.7 & -- \\
 & $\zeta(6f)$ & 21.5 & 2.1 & 0.098 \\
 4f$^{13}$5d$^{2}$ & $E_{\rm av}$ & 143\,806.2 & 160\,620.6 & -- \\
 & $F^{2}(5d,5d)$ & 59\,915.6 & 46\,172.0 & 0.771 \\
 & $F^{4}(5d,5d)$ & 39\,820.9 & 22\,252.0 & 0.559 \\
 & $\zeta(4f)$ & 3\,879.1 & 3\,853.0 & 0.993 \\
 & $\zeta(5d)$ & 2\,844.9 & 2\,298.0 & 0.808 \\
 & $F^{2}(4f,5d)$ & 35\,718.4 & 25\,801.0 & 0.722 \\
 & $F^{4}(4f,5d)$ & 17\,124.8 & 14\,208.0 & 0.830 \\
 & $G^{1}(4f,5d)$ & 14\,215.2 & 9\,495.0 & 0.668 \\
 & $G^{3}(4f,5d)$ & 12\,342.2 & 12\,856.0 & 1.042 \\
 & $G^{5}(4f,5d)$ & 9\,618.4 & 7\,842.0 & 0.815 \\
 4f$^{14}$5f - 4f$^{13}$5d$^{2}$ & $R^{1}$ & $-8\,553.8$ & $-6\,345.7$ & 0.742 \\
 & $R^{3}$ & $-4\,353.5$ & $-3\,083.8$ & 0.708 \\
 & $R^{5}$ & $-2\,691.7$ & $-1\,860.2$ & 0.691 \\
 4f$^{14}$5d & $E_{\rm av}$ & 3\,907.9 & 2\,900.5 & -- \\
 & $\zeta(5d)$ & 2\,548.0 & 1\,876.7 & 0.737 \\
 4f$^{14}$6d & $E_{\rm av}$ & 140\,552.8 & 141\,663.1 & -- \\
 & $\zeta(6d)$ & 502.2 & 522.3 & 1.040 \\
 4f$^{14}$6s & $E_{\rm av}$ & 20\,615.8 & 19\,281.4 & -- \\
 4f$^{14}$7s & $E_{\rm av}$ & 139\,907.8 & 140\,270.9 & -- \\
 4f$^{14}$8s & $E_{\rm av}$ & 188\,717.7 & 189\,893.6 & -- \\
\end{longtable}
\noindent\footnotesize The HFR column gives the ab initio parameter values and the HFR+fit column the fitted values. Ratio = (HFR+fit)/HFR. Ratios are not quoted for $E_{\rm av}$ or for effective parameters whose HFR value is zero.
\endgroup

\begingroup
\scriptsize
\setlength{\tabcolsep}{3.5pt}
\begin{longtable}{@{}llrrr@{}}
\caption{HFR and HFR+fit parameters for \ion{Tl}{iv}.}\label{tab:tl-fit-params}\\
\toprule
Configuration & Parameter & HFR (cm$^{-1}$) & HFR+fit (cm$^{-1}$) & Ratio \\
\midrule
\endfirsthead
\multicolumn{5}{c}{\tablename~\thetable\ (continued)}\\
\toprule
Configuration & Parameter & HFR (cm$^{-1}$) & HFR+fit (cm$^{-1}$) & Ratio \\
\midrule
\endhead
\midrule \multicolumn{5}{r}{Continued on next page}\\ \endfoot
\bottomrule \endlastfoot
 5d$^{10}$ & $E_{\rm av}$ & 2\,367.9 & 2\,378.3 & -- \\
 5d$^{9}$6s & $E_{\rm av}$ & 89\,728.1 & 86\,553.5 & -- \\
 & $\zeta(5d)$ & 7\,526.7 & 7\,385.5 & 0.981 \\
 & $G^{2}(5d,6s)$ & 21\,822.9 & 18\,702.8 & 0.857 \\
 5d$^{9}$7s & $E_{\rm av}$ & 263\,161.6 & 263\,783.4 & -- \\
 & $\zeta(5d)$ & 7\,661.7 & 7\,521.7 & 0.982 \\
 & $G^{2}(5d,7s)$ & 3\,735.5 & 2\,975.3 & 0.796 \\
 5d$^{9}$6d & $E_{\rm av}$ & 262\,590.8 & 264\,769.0 & -- \\
 & $\zeta(5d)$ & 7\,669.6 & 7\,515.5 & 0.980 \\
 & $\zeta(6d)$ & 679.4 & 842.5 & 1.240 \\
 & $F^{2}(5d,6d)$ & 13\,990.6 & 12\,181.4 & 0.871 \\
 & $F^{4}(5d,6d)$ & 5\,904.8 & 5\,604.3 & 0.949 \\
 & $G^{0}(5d,6d)$ & 2\,843.9 & 2\,342.2 & 0.824 \\
 & $G^{2}(5d,6d)$ & 3\,675.8 & 3\,618.0 & 0.984 \\
 & $G^{4}(5d,6d)$ & 3\,048.5 & 2\,598.2 & 0.852 \\
 5d$^{9}$6p & $E_{\rm av}$ & 173\,963.5 & 173\,032.7 & -- \\
 & $\zeta(5d)$ & 7\,596.4 & 7\,381.4 & 0.972 \\
 & $\zeta(6p)$ & 12\,101.9 & 13\,181.6 & 1.089 \\
 & $F^{2}(5d,6p)$ & 29\,216.0 & 26\,714.7 & 0.914 \\
 & $G^{1}(5d,6p)$ & 11\,084.5 & 10\,268.7 & 0.926 \\
 & $G^{3}(5d,6p)$ & 9\,568.0 & 10\,500.0 & 1.097 \\
 5d$^{9}$7s - 5d$^{9}$6d & $R^{2}(5d\,7s;5d\,6d)$ & $-3\,471.5$ & $-3\,098.2$ & 0.892 \\
 & $R^{2}(5d\,7s;6d\,5d)$ & $-94.6$ & $-84.5$ & 0.892 \\
\end{longtable}
\noindent\footnotesize The HFR column gives the ab initio parameter values and the HFR+fit column the fitted values. Ratio = (HFR+fit)/HFR. Ratios are not quoted for $E_{\rm av}$ or for effective parameters whose HFR value is zero.
\endgroup

\newpage 

\section{Energy Levels}
\label{app:levels}
This appendix collects the experimental and calculated energy levels used in the present work for \ion{As}{III}, \ion{Se}{III}, \ion{Hf}{IV}, and \ion{Tl}{IV}. The tables give the experimental energies and the corresponding HFR+fit energies, together with their absolute relative deviations from experiment. MCDHF energies and their corresponding deviations are also reported where available. The level indices are used to identify the lower and upper states in the radiative-transition tables of Appendix~C. 

\begingroup
\scriptsize
\setlength{\tabcolsep}{3.5pt}
\begin{longtable}{@{}rlllrrrrr@{}}
\caption{Experimental and HFR+fit energy levels used in the least-squares optimization of \ion{As}{iii}. Energies are in cm$^{-1}$.}\label{tab:as-fit-levels}\\
\toprule
Index & Configuration & Term & $J$ & $E_{\rm exp}$ & $E_{\rm HFR+fit}$ & $\Delta_{\rm HFR+fit}$ (\%) & $E_{\rm MCDHF}$ & $\Delta_{\rm MCDHF}$ (\%) \\
\midrule
\endfirsthead
\multicolumn{7}{c}{\tablename~\thetable\ (continued)}\\
\toprule
Index & Configuration & Term & $J$ & $E_{\rm exp}$ & $E_{\rm HFR+fit}$ & $\Delta_{\rm HFR+fit}$ (\%) & $E_{\rm MCDHF}$ & $\Delta_{\rm MCDHF}$ (\%) \\
\midrule
\endhead
\midrule \multicolumn{7}{r}{Continued on next page}\\ \endfoot
\bottomrule \endlastfoot
1 & 4s$^{2}$4p & $^{2}\mathrm{P}$ & 1/2 & 0.0 & 0 & --                    & 0.0 & -- \\
2 & 4s$^{2}$4p & $^{2}\mathrm{P}$ & 3/2 & 2\,940.0 & 2\,940 & 0.0000      & 3\,150   & 7.14 \\
3 & 4s4p$^{2}$ & $^{2}\mathrm{D}$ & 3/2 & 85\,310.0 & 85\,349 & 0.046     & 81\,492  & 4.48\\
4 & 4s4p$^{2}$ & $^{2}\mathrm{D}$ & 5/2 & 85\,633.0 & 85\,577 & 0.065     & 81\,826  & 4.45\\
5 & 4s$^{2}$5s & $^{2}\mathrm{S}$ & 1/2 & 106\,694.0 & 106\,700 & 0.006   & 103\,660 & 2.84\\
6 & 4s4p$^{2}$ & $^{2}\mathrm{S}$ & 1/2 & 107\,808.0 & 107\,809 & 0.0009  & 105\,013 & 2.59\\
7 & 4s4p$^{2}$ & $^{2}\mathrm{P}$ & 1/2 & 113\,936.0 & 113\,925 & 0.010   & 111\,863 & 1.82\\
8 & 4s4p$^{2}$ & $^{2}\mathrm{P}$ & 3/2 & 115\,424.0 & 115\,430 & 0.005   & 113\,318 & 1.82\\
\end{longtable}
\endgroup

\begingroup
\scriptsize
\setlength{\tabcolsep}{3.5pt}
\begin{longtable}{@{}rlllrrr@{}}
\caption{Experimental and HFR+fit energy levels used in the least-squares optimization of \ion{Se}{iii}. Energies are in cm$^{-1}$.}\label{tab:se-fit-levels}\\
\toprule
Index & Configuration & Term & $J$ & $E_{\rm exp}$ & $E_{\rm HFR+fit}$ & $\Delta_{\rm HFR+fit}$ (\%) \\
\midrule
\endfirsthead
\multicolumn{7}{c}{\tablename~\thetable\ (continued)}\\
\toprule
Index & Configuration & Term & $J$ & $E_{\rm exp}$ & $E_{\rm HFR+fit}$ & $\Delta_{\rm HFR+fit}$ (\%) \\
\midrule
\endhead
\midrule \multicolumn{7}{r}{Continued on next page}\\ \endfoot
\bottomrule \endlastfoot
1 & 4s$^{2}$4p$^{2}$ & $^{3}\mathrm{P}$ & 0 & 0.0 & 0 & -- \\
2 & 4s$^{2}$4p$^{2}$ & $^{3}\mathrm{P}$ & 1 & 1\,742.4 & 1\,757 & 0.838 \\
3 & 4s$^{2}$4p$^{2}$ & $^{3}\mathrm{P}$ & 2 & 3\,936.0 & 3\,935 & 0.025 \\
4 & 4s$^{2}$4p$^{2}$ & $^{1}\mathrm{D}$ & 2 & 13\,032.0 & 13\,037 & 0.038 \\
5 & 4s$^{2}$4p$^{2}$ & $^{1}\mathrm{S}$ & 0 & 28\,128.7 & 28\,133 & 0.015 \\
6 & 4s4p$^{3}$ & $^{5}\mathrm{S}$ & 2 & 68\,502.7 & 68\,499 & 0.005 \\
7 & 4s4p$^{3}$ & $^{3}\mathrm{D}$ & 1 & 92\,611.2 & 92\,666 & 0.059 \\
8 & 4s4p$^{3}$ & $^{3}\mathrm{D}$ & 2 & 92\,726.3 & 92\,721 & 0.006 \\
\end{longtable}
\endgroup

\begingroup
\scriptsize
\setlength{\tabcolsep}{3.5pt}
\begin{longtable}{@{}rlllrrrrr@{}}
\caption{Experimental and HFR+fit energy levels used in the least-squares optimization of \ion{Hf}{iv}. Energies are in cm$^{-1}$.}\label{tab:hf-fit-levels}\\
\toprule
Index & Configuration & Term & $J$ & $E_{\rm exp}$ & $E_{\rm HFR+fit}$ & $\Delta_{\rm HFR+fit}$ (\%) & $E_{\rm MCDHF}$ & $\Delta_{\rm MCDHF}$ (\%) \\
\midrule
\endfirsthead
\multicolumn{9}{c}{\tablename~\thetable\ (continued)}\\
\toprule
Index & Configuration & Term & $J$ & $E_{\rm exp}$ & $E_{\rm HFR+fit}$ & $\Delta_{\rm HFR+fit}$ (\%) & $E_{\rm MCDHF}$ & $\Delta_{\rm MCDHF}$ (\%) \\
\midrule
\endhead
\midrule \multicolumn{9}{r}{Continued on next page}\\ \endfoot
\bottomrule \endlastfoot
1 & 4f$^{14}$5d & $^{2}\mathrm{D}$ & 1.5 & 0.0 & 0 & --                    & 0 & -- \\
2 & 4f$^{14}$5d & $^{2}\mathrm{D}$ & 2.5 & 4\,692.6 & 4\,693 & 0.009       & 4\,478 & 4.57 \\
3 & 4f$^{14}$6s & $^{2}\mathrm{S}$ & 0.5 & 18\,380.1 & 18\,380 & 0.0005    & 19\,106 & 3.95  \\
4 & 4f$^{14}$6p & $^{2}\mathrm{P}$ & 0.5 & 67\,039.0 & 67\,040 & 0.001     & 76\,131 & 13.56 \\
5 & 4f$^{14}$6p & $^{2}\mathrm{P}$ & 1.5 & 76\,614.2 & 76\,615 & 0.001     & 84\,877 & 10.78 \\
6 & 4f$^{14}$7s & $^{2}\mathrm{S}$ & 0.5 & 140\,226.2 & 140\,226 & 0.0001  & 140\,599 & 0.27 \\
7 & 4f$^{14}$6d & $^{2}\mathrm{D}$ & 1.5 & 140\,709.3 & 140\,709 & 0.0002  & 141\,065 & 0.25 \\
8 & 4f$^{14}$6d & $^{2}\mathrm{D}$ & 2.5 & 142\,024.1 & 142\,024 & 0.0001  & 142\,350 & 0.23 \\
\end{longtable}
\endgroup

\begingroup
\scriptsize
\setlength{\tabcolsep}{3.5pt}
\begin{longtable}{@{}rlllrrrrr@{}}
\caption{Experimental and HFR+fit energy levels used in the least-squares optimization of \ion{Tl}{iv}. Energies are in cm$^{-1}$.}\label{tab:tl-fit-levels}\\
\toprule
Index & Configuration & Term & $J$ & $E_{\rm exp}$ & $E_{\rm HFR+fit}$ & $\Delta_{\rm HFR+fit}$ (\%) & $E_{\rm MCDHF}$ & $\Delta_{\rm MCDHF}$ (\%) \\
\midrule
\endfirsthead
\multicolumn{9}{c}{\tablename~\thetable\ (continued)}\\
\toprule
Index & Configuration & Term & $J$ & $E_{\rm exp}$ & $E_{\rm HFR+fit}$ & $\Delta_{\rm HFR+fit}$ (\%) & $E_{\rm MCDHF}$ & $\Delta_{\rm MCDHF}$ (\%) \\
\midrule
\endhead
\midrule \multicolumn{9}{r}{Continued on next page}\\ \endfoot
\bottomrule \endlastfoot
1 & 5d$^{10}$ & $^{1}\mathrm{S}$ & 0 & 0.0 & 0 & -- &0 & --\\
2 & 5d$^{9}$6s & $^{3}\mathrm{D}$ & 3 & 75\,049.6 & 75\,152 & 0.136     & 79\,091  & 5.38\\
3 & 5d$^{9}$6s & $^{3}\mathrm{D}$ & 2 & 78\,644.0 & 78\,536 & 0.137     & 82\,870  & 5.37\\
4 & 5d$^{9}$6s & $^{3}\mathrm{D}$ & 1 & 93\,674.8 & 93\,602 & 0.078     & 97\,388  & 3.96\\
5 & 5d$^{9}$6s & $^{1}\mathrm{D}$ & 2 & 96\,725.1 & 96\,804 & 0.082     & 100\,742 &  4.15\\
6 & 5d$^{9}$6p & $^{3}\mathrm{P}$ & 2 & 147\,632.4 & 147\,842 & 0.142   & 151\,459 & 2.59 \\
7 & 5d$^{9}$6p & $^{3}\mathrm{F}$ & 3 & 149\,837.4 & 149\,937 & 0.066   & 153\,663 & 2.55 \\
8 & 5d$^{9}$6p & $^{3}\mathrm{F}$ & 2 & 166\,421.6 & 166\,248 & 0.104   & 169\,858 & 2.06 \\
\end{longtable}
\endgroup


\section{Radiative transitions}
\label{app:transitions}
This appendix collects the radiative transition data obtained for \ion{As}{iii}, \ion{Se}{iii}, \ion{Hf}{iv}, and \ion{Tl}{iv}. The tables give the experimental wavelengths when available, the Ritz wavelengths, the indices of the lower and upper levels defined in Appendix~B, and the HFR oscillator strengths. The corresponding MCDHF oscillator strengths are also reported for the transitions for which they are available.

\begingroup
\scriptsize
\setlength{\tabcolsep}{3.5pt}
\begin{longtable}{@{}rrcrr@{}}
\caption{Radiative transitions of \ion{As}{iii} calculated with HFR. Wavelengths are in \AA.}\label{tab:as-hfrcpol-transitions}\\
\toprule
$\lambda_{\rm exp}$ (\AA) & $\lambda_{\rm Ritz}$ (\AA) & Transition & $\log(gf)_{\rm HFR+fit}$ & $\log(gf)_{\rm MCDHF}$ \\
\midrule
\endfirsthead
\multicolumn{5}{c}{\tablename~\thetable\ (continued)}\\
\toprule
$\lambda_{\rm exp}$ (\AA) & $\lambda_{\rm Ritz}$ (\AA) & Transition & $\log(gf)_{\rm HFR+fit}$ & $\log(gf)_{\rm MCDHF}$ \\
\midrule
\endhead
\midrule \multicolumn{5}{r}{Continued on next page}\\ \endfoot
\bottomrule \endlastfoot
603.81 & 603.78 & 1--18 & -0.654 &    -0.674 \\
-- & 613.91 & 1--15 & -1.330 &        -1.433 \\
614.41 & 614.37 & 2--19 & -0.404 &    -0.434 \\
614.73 & 614.69 & 2--18 & -1.338 &    -1.376 \\
-- & 625.19 & 2--15 & -1.035 &        -1.100 \\
849.94 & 849.97 & 1--9 & 0.264 &      0.173  \\
871.04 & 871.05 & 2--10 & 0.668 &     0.666  \\
871.75 & 871.76 & 2--9 & 0.131 &      0.221  \\
\end{longtable}
\endgroup

\begingroup
\scriptsize
\setlength{\tabcolsep}{3.5pt}
\begin{longtable}{@{}rrcr@{}}
\caption{Radiative transitions of \ion{Se}{iii} calculated with HFR. Wavelengths are in \AA.}\label{tab:se-hfrcpol-transitions}\\
\toprule
$\lambda_{\rm exp}$ (\AA) & $\lambda_{\rm Ritz}$ (\AA) & Transition & $\log(gf)_{\rm HFR+fit}$ \\
\midrule
\endfirsthead
\multicolumn{4}{c}{\tablename~\thetable\ (continued)}\\
\toprule
$\lambda_{\rm exp}$ (\AA) & $\lambda_{\rm Ritz}$ (\AA) & Transition & $\log(gf)_{\rm HFR+fit}$ \\
\midrule
\endhead
\midrule \multicolumn{4}{r}{Continued on next page}\\ \endfoot
\bottomrule \endlastfoot
430.560 & 430.56 & 1--85 & -2.443 \\
-- & 431.71 & 1--82 & -4.378 \\
-- & 432.86 & 1--78 & -3.250 \\
-- & 433.82 & 2--85 & -2.829 \\
434.913 & 434.91 & 2--83 & -2.117 \\
434.980 & 434.98 & 2--82 & -1.777 \\
435.070 & 435.07 & 2--81 & -1.833 \\
435.618 & 435.62 & 2--79 & -2.059 \\
\end{longtable}
\endgroup

\begingroup
\scriptsize
\setlength{\tabcolsep}{3.5pt}
\begin{longtable}{@{}rrcrr@{}}
\caption{Radiative transitions of \ion{Hf}{iv} calculated with HFR. Wavelengths are in \AA.}\label{tab:hf-hfrcpol-transitions}\\
\toprule
$\lambda_{\rm exp}$ (\AA) & $\lambda_{\rm Ritz}$ (\AA) & Transition & $\log(gf)_{\rm HFR+fit}$ & $\log(gf)_{\rm MCDHF}$ \\
\midrule
\endfirsthead
\multicolumn{5}{c}{\tablename~\thetable\ (continued)}\\
\toprule
$\lambda_{\rm exp}$ (\AA) & $\lambda_{\rm Ritz}$ (\AA) & Transition & $\log(gf)_{\rm HFR+fit}$ & $\log(gf)_{\rm MCDHF}$ \\
\midrule
\endhead
\midrule \multicolumn{5}{r}{Continued on next page}\\ \endfoot
\bottomrule \endlastfoot
508.111 & 508.11 & 1--37 & -0.253 & -0.373 \\
520.039 & 520.04 & 2--38 & -0.080 & -0.194 \\
520.518 & 520.52 & 2--37 & -1.332 & -1.498\\
579.825 & 579.82 & 1--35 & -2.458 & -- \\
581.77 & 581.77 & 1--34 & -0.660 & -- \\
582.01 & 582.01 & 1--33 & -2.797 & -- \\
-- & 593.42 & 1--30 & -0.880 & -- \\
596.561 & 596.56 & 1--28 & -0.617 & -- \\
\end{longtable}
\endgroup

\begingroup
\scriptsize
\setlength{\tabcolsep}{3.5pt}
\begin{longtable}{@{}rrcrr@{}}
\caption{Radiative transitions of \ion{Tl}{iv} calculated with HFR. Wavelengths are in \AA.}\label{tab:tl-hfrcpol-transitions}\\
\toprule
$\lambda_{\rm exp}$ (\AA) & $\lambda_{\rm Ritz}$ (\AA) & Transition & $\log(gf)_{\rm HFR+fit}$ & $\log(gf)_{\rm MCDHF}$ \\
\midrule
\endfirsthead
\multicolumn{5}{c}{\tablename~\thetable\ (continued)}\\
\toprule
$\lambda_{\rm exp}$ (\AA) & $\lambda_{\rm Ritz}$ (\AA) & Transition & $\log(gf)_{\rm HFR+fit}$ & $\log(gf)_{\rm MCDHF}$ \\
\midrule
\endhead
\midrule \multicolumn{5}{r}{Continued on next page}\\ \endfoot
\bottomrule \endlastfoot
531.257 & 531.26 & 1--16 & -0.894 & -0.839 \\
570.488 & 570.50 & 1--13 & -0.187 &  0.006\\
597.024 & 597.02 & 1--9 & -2.281 & -1.740 \\
-- & 773.79 & 6--38 & -2.221 & -3.170 \\
-- & 778.03 & 6--37 & -2.471 & -3.220 \\
-- & 784.14 & 6--35 & -2.774 &  -1.120\\
-- & 786.55 & 6--34 & -3.242 & -3.640 \\
-- & 787.23 & 7--38 & -2.520 &  -2.960\\
\end{longtable}
\endgroup


\begin{adjustwidth}{-\extralength}{0cm}

\reftitle{References}


 \bibliography{hst_heavy}

\PublishersNote{}
\end{adjustwidth}
\end{document}